\documentclass[aip,
 amsmath,amssymb,
 reprint,%
floatfix,
]{revtex4-1}
\usepackage{graphicx}
\usepackage{dcolumn}
\usepackage{bm}
\usepackage[mathlines]{lineno}
\usepackage[utf8]{inputenc}
\usepackage[T1]{fontenc}
\DeclareUnicodeCharacter{0394}{$\Delta$}
\usepackage{mathptmx}
\usepackage{etoolbox}
\usepackage[table]{xcolor}
\usepackage{booktabs}
\usepackage{url}
\usepackage{hyperref}
\usepackage{multirow}
\usepackage{tabularx}
\newcolumntype{Y}{>{\centering\arraybackslash}X}

\makeatletter
\def\@email#1#2{%
 \endgroup
 \patchcmd{\titleblock@produce}
  {\frontmatter@RRAPformat}
  {\frontmatter@RRAPformat{\produce@RRAP{*#1\href{mailto:#2}{#2}}}\frontmatter@RRAPformat}
  {}{}
}%
\makeatother
\makeatletter
\let\selectlanguage\@gobble
\makeatother
\begin{document}
\preprint{AIP/123-QED}
\title{Multi-Task Graph Neural Network Predictions of Auger-Electron and X-ray Photoelectron Spectroscopy}
\author{Adam E. A. Fouda*}
\affiliation{Department of Physics, The University of Chicago, Chicago, IL 60637, USA}
\email{adamfouda@uchicago.edu}
\affiliation{Chemical Sciences and Engineering Division, Argonne National Laboratory, 9700 S. Cass Avenue, Lemont, IL 60439, USA}
\author{Patrick Phillips}
\affiliation{Chemical Sciences and Engineering Division, Argonne National Laboratory, 9700 S. Cass Avenue, Lemont, IL 60439, USA}
\affiliation{Department of Computer Science, The University of Chicago, Chicago, Illinois 60637, United States}
\author{Phay J. Ho}
\affiliation{Chemical Sciences and Engineering Division, Argonne National Laboratory, 9700 S. Cass Avenue, Lemont, IL 60439, USA}

\begin{abstract}
Auger-electron spectroscopy has long accompanied x-ray photoelectron spectroscopy as a second modality to resolve chemical states with overlapping core-electron binding energies. However, analyzing the Auger spectrum is challenged by its complexity and the computational expense of its simulation. Here we demonstrate that the physical connection, and thus inter-task relationship, between the generation of a core-hole and its corresponding Auger-Meitner decay enables inductive knowledge transfer through the training of a multi-task graph neural network to predict both observables from a common graph embedding. Both task losses are combined with learned weights via the uncertainty weighting procedure. Overall, the single-task and multi-task models predict calculated and experimental Auger lineshapes with good accuracy in most cases. The performance between the two task regimes is similar, with the single-task models generally having better predictions of the finer peak structures in the spectrum. The present results demonstrate that multi-task training is a promising avenue for future developments of universal x-ray spectroscopy models with learned representations that map the molecular structure to a multitude of techniques.
\end{abstract}
\maketitle

\section{Introduction}

Photoelectrons emitted from chemical samples following x-ray irradiation are affected by the local bond environments of the target atoms. This enables x-ray photoelectron spectroscopy (XPS) to identify different chemical states among common atom types in materials,\cite{bagus_interpretation_2013} molecules,\cite{siegbahn_esca_1970} solutions,\cite{seidel_valence_2016} liquids,\cite{hurisso_amino_2011,dick_probing_2020} and biological matter.\cite{ratner_surface_1983} While this phenomenon, known as the chemical shift,\cite{bagus_mechanisms_1999} can resolve different oxidation states and local bond environments within the XPS spectrum, overlapping peaks in the spectrum can limit this capability through XPS measurements alone. Moreover, the availability of reliable reference spectra to support precise peak assignments is challenged by the influence of multiple competing mechanisms, such as charge transfer, electric fields and hybridization,\cite{bagus_mechanisms_1999} which govern the structural environment effects on the binding energies of core-electrons.

Combining multiple complementary spectral measurements yields a more holistic observation of the chemical system and can resolve the spectral ambiguities of a single measurement. XPS has long been implemented with the simultaneous detection of electrons autoionized from the irradiated sample via the core-hole decay of the ionized atom.\cite{Siegbahn1967ESCAA,10.1021/ac60314a015} This decay process, known as Auger-Meitner decay, involves an outer-shell electron filling the core vacancy, and
another outer-shell electron being ejected to the continuum. Auger-electron spectroscopy (AES) is often used to distinguish oxidation states of active metal sites in catalysts and functional materials.\cite{doi:10.1021/acsanm.5c00100,https://doi.org/10.1002/sia.6239,SOLDEMO2024122565,HENDERSON2025147578,FOX1977390,10.1039/c2cp22419d,https://doi.org/10.1002/admi.202201828,doi:10.1021/jp0564400,doi:10.1021/acsomega.1c05002} This multi-modal analysis is primarily implemented via a Wagner plot,\cite{MORETTI20133} which plots the XPS core-electron binding energy ($E_{b}$) against Auger kinetic energy ($E_{k}$). The Wagner plot is thus a representation of the Auger parameter ($\alpha$),\cite{WAGNER1988283}
\begin{equation}\label{eq:alpha}
\alpha = E_{b} + E_{k},
\end{equation}
which assumes that the binding energies of electrons resulting in $E_{k}$ are deep core orbitals, and therefore approximates them to have equal couplings to the valence orbitals. This approximation enables the Auger-parameter and Wagner plot to isolate different final state relaxation effects in the Auger-Meitner decay, which are more sensitive to correlation effects,\cite{EGELHOFF1987253} and thus resolve different chemical states with overlapping $E_{b}$ values.

Beyond its complementary chemical state identification capabilities to XPS, AES is a common observable for fundamental molecular physics applications exploring novel interactions under ultra-fast and intense x-ray free electron laser conditions\cite{Pelimanni2024} and monitoring ultra-fast dynamics in time-resolved and imaging studies.\cite{Thompson_2024,D1CP00623A,doi:10.1126/science.abj2096,Driver2024,Simmermacher2026} AES is a powerful observable for light-induced phenomena with relatively low cross-sections in organic molecules, as the underlying decay mechanism out-competes fluorescence decay in the first row $p$-block elements. Furthermore, Auger processes play a key role in the formation of radiation damage effects that influence structural x-ray diffraction measurements, and are utilized in medical radiotherapy applications.\cite{doi:10.1080/09553002.2020.1831706,Ku2019,Stollenwerk_2025}

\begin{figure}[!htbp]
    \centering
    \includegraphics[width=1.0\columnwidth]{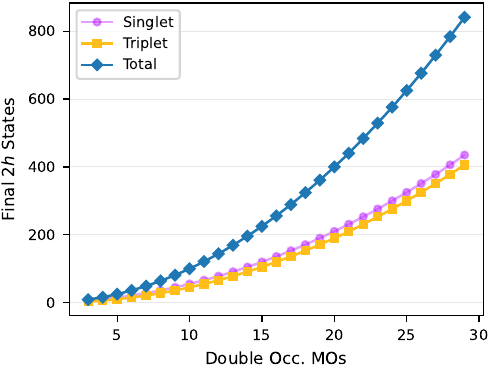}
    \caption{Plot showing how the number of singlet and triplet final dication states (excluding multielectron shakeup and shakeoff processes) increases with respect to the number of doubly occupied orbitals above the core-ionized orbital of a closed-shell ground-state system by Weyl's formula in Equation \ref{eq:weyl}.}
    \label{fig:augerstates}
\end{figure}

The implementation of AES analysis to the aforementioned applications is limited by the complexity of Auger-Meitner decay. All energetically possible decay channels involving two outer-shell electrons contribute to the spectrum, which challenges both its experimental interpretation and computational simulation. The number of possible final states scales non-linearly with respect to the system size. This can be illustrated in the case of a closed shell ground state system undergoing core-ionization. By excluding the contribution of multielectron shakeup and shakeoff processes, the number of final states can be approximated by Weyl’s formula for determining the number of configuration state functions in an active orbital space,
\begin{equation}\label{eq:weyl}
K(n,N,S) = \frac{2S+1}{n+1}\binom{n+1}{\frac{1}{2}N-S}\binom{n+1}{\frac{1}{2}N+S+1},
\end{equation}
where $n$, $N$, and $S$ are the number of molecular orbitals, the number of electrons, and the spin quantum number, respectively. Figure \ref{fig:augerstates} uses Equation \ref{eq:weyl} to plot the number of singlet and triplet final dication states, with respect to the number of doubly occupied orbitals energetically higher than the core-ionized orbital. The computation of these states is further complicated by their multiconfigurational character,\cite{doi:10.1063/1.1386414,doi:10.1021/acs.jpca.5c01789} which limits the capability of single reference electronic structure methods, such as density functional theory (DFT), to describe the states accurately. Another complexity in AES simulation is the treatment of the ejected electron's continuum wave function and its interaction with the dication molecular frame. Numerous approaches have been developed for its treatment, including approximating the spectral intensity by an electron population analysis,\cite{MITANI2003103,10.1063/1.2166234,D3CP01746J,doi:10.1021/acs.jpclett.3c03611} and both implicit\cite{10.1063/1.1316046,BSchimmelpfennig_1992,SCHIMMELPFENNIG1995173,LIEGENER1982188,10.1063/1.2126976,10.1063/5.0036976,SIEGBAHN1975330,JENNISON1980435,Larkins1990,FINK1995295,TRAVNIKOVA200967,PhysRevA.94.023422,10.1063/1.4919794} and explicit\cite{PhysRevA.19.1649,HIGASHI1982377,PhysRevA.45.318,Demekhin2007,PhysRevA.80.063425,10.1063/1.3526026,10.1063/1.3700233,C7CP02345F} considerations of the continuum electron wave function.

Previously, we found that implicitly treating the continuum with pre-calculated bound-continuum integrals from atomic calculations via the one center approximation (OCA)\cite{SIEGBAHN1975330,JENNISON1980435,Larkins1990,FINK1995295,TRAVNIKOVA200967,PhysRevA.94.023422,10.1063/1.4919794} implementation in OpenMolcas,\cite{tenorio_multi-reference_2021,fdez2019openmolcas} combined with the multiconfiguration pair-density functional theory\cite{mcpdft2014,mcpdftreview01, mcpdftreview02, mcpdftreview03} (MC-PDFT) method for treating the bound electronic structure, efficiently provides carbon 1$s$ Auger spectra for a set of 20 organic molecules with reasonable experiment accuracy.\cite{doi:10.1021/acs.jpca.5c01789} MC-PDFT enables an efficient description of the large number of doubly ionized final states in AES. It adds correlation effects to a restricted active space self-consistent field wave function\cite{werner1981quadratically,malmqvist1990restricted} (RASSCF), through a straightforward computation of the one- and two-particle reduced density matrices, and the optimized orbitals derived from the RASSCF wave function. The RASSCF protocol enables both the description of core-hole and multiconfigurational valence dication states. Traditionally, the calculation of the RASSCF wave function was followed by second order perturbation theory (RASPT2) to include additional effects of the correlation energy.\cite{malmqvist2008restricted} However, the RASPT2 method requires the calculation of higher-order density matrices to determine the perturbed wave function used to compute the energy. MC-PDFT has been shown to achieve an accuracy comparable to RASPT2 for the energies and properties of the ground and excited states,\cite{mcpdftreview02, mcpdftreview03} and more recently AES.\cite{doi:10.1021/acs.jpca.5c01789} The OCA MC-PDFT level of theory thus enables the possibility to construct high-quality AES datasets.

Recently, there has been a widespread surge in the development of data driven, machine learning (ML) approaches to predicting spectroscopy.\cite{rankine_progress_2021,10.1039/d5sc05628d} The popularity of such methods is driven by their ability to reproduce calculated spectra at comparatively negligible computational cost. To the best of our knowledge, no previous ML models for predicting molecular Auger spectra directly from structure have been reported. However, neural networks have been applied to the spectrum-to-structure analysis of AES\cite{https://doi.org/10.1002/sia.740201303,https://doi.org/10.1002/sia.740230709,https://doi.org/10.1002/sia.5011} and Budewig \textit{et al.} deep neural networks to the prediction of multiple x-ray induced transitions, including Auger-Meitner decay rates, in atomic systems.\cite{PhysRevResearch.6.013265} Furthermore, multiple deep learning models have been developed for the prediction of x-ray absorption spectroscopy (XAS)\cite{rankine_deep_2020,rankine_accurate_2022,madkhali_role_2020,carbone_machine-learning_2020,kotobi_integrating_2023-1,kharel_omnixas_2025,luder_machine_2025,zhan_graph_2025,gleason_cuxasnet_2025,prange_toward_2025} and XPS,\cite{sun_machine_2022,bartok_representing_2013,mejia-rodriguez_scalable_2021,mejia-rodriguez_basis_2022,golze_accurate_2022,zarrouk_experiment-driven_2024,tripathy_chemical_2024,porcelli_photoemission_2025} due to the availability of databases\cite{CIBIN2020108479,Guo2023} and high-throughput simulation methods.\cite{Mathew2018,10.1039/b926434e,10.1021/acs.jpcc.4c03339} Graph neural networks (GNNs) are a class of deep geometric learning methods, which have gained broad popularity in the chemical sciences,\cite{buterez_transfer_2024,schutt_equivariant_nodate,xu_pretrained_2025} including the prediction of XAS,\cite{kotobi_integrating_2023-1,zhan_graph_2025,gleason_cuxasnet_2025} as they embed the molecular structure directly into the architecture via the representation of atoms and bonds as the network nodes and edges. This alleviates the requirement to tune appropriate input descriptors to numerically represent the local bond environment in non-geometric model architectures. Furthermore, GNN models can incorporate 3D geometric information by respecting the invariant or equivariant nature of molecular properties to rotational and translational symmetries.\cite{satorras_en_2022,schutt_equivariant_nodate,batatia_mace_2023,geiger_e3nn_2022-1,passaro_reducing_2023} Recently, we have shown that the equivariant graph neural network (EGNN), which includes 3D geometry effects via an equivariant coordinate update message operation,\cite{satorras_en_2022} accurately predicts XPS $E_{b}$ values for organic molecules with respect to experiment, when trained on values calculated by MC-PDFT.\cite{fouda_2026_19689244}

Here, we extend this EGNN framework to direct prediction of molecular C 1s Auger spectra from molecular structure. We construct a large OCA/MC-PDFT AES dataset and investigate whether the physical relationship between XPS and AES can be exploited through multi-task learning of $E_b$ and $f$($E_k$). We compare the predictive performance and generalization of single- and multi-task models for both observables and show that both regimes  reproduce the overall lineshape with a similar level of accuracy. The work demonstrates the applicability of multi-task learning schemes to x-ray spectroscopy, where the observables from different techniques share are common relationship to the local-bond environment.

\begin{figure}[!htbp]
    \centering
    \includegraphics[width=1.0\columnwidth]{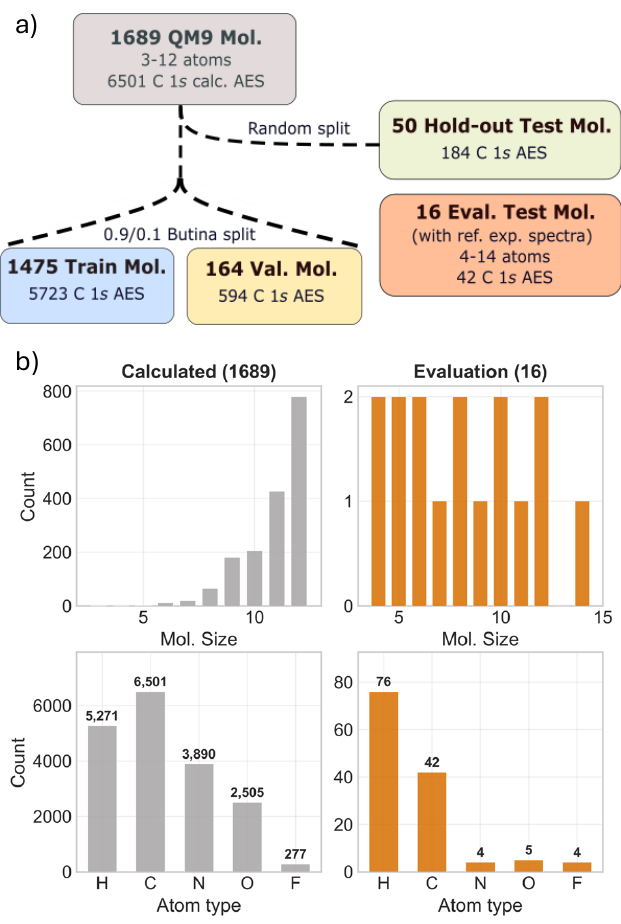}
    \caption{a) Schematic showing the split of the Auger spectra calculated on the subset of QM9 molecules (grey) between the train (blue), validation (yellow) and hold-out test (green) sets, and the additional set of evaluation molecules with experimental and calculated spectra (orange). b) Plots of the molecular size (top row) and atom types (bottom row) counts of the QM9 calculated dataset (left column, grey) and the evaluation dataset (right column, orange).}
    \label{fig:data}
\end{figure}

\section{Dataset and Model}

This work was enabled through the construction of an AES dataset by the previously benchmarked OCA MC-PDFT method with the tPBE0 functional and ANO-RCC-VTZP basis set. The approach uses a RASSCF active space that excludes virtual ground state orbitals and thus enables high-throughput AES simulation.\cite{doi:10.1021/acs.jpca.5c01789}  We refer the reader to this study for additional details on the electronic structure calculations. The AES dataset contains 1689 molecules, containing 3-12 atoms, extracted from a subset of randomly selected QM9 molecules\cite{ramakrishnan_quantum_2014} previously used for EGNN predictions of carbon 1$s$ $E_{b}$ values.\cite{fouda2026experimentallyaccurategraphneural} The number of final dication states for each molecule was limited to 300 singlet and 300 triplet states and the Auger spectra for every possible carbon $1s$ hole state to these final states were computed. The singlet and triplet final states for each carbon were combined and fitted to a kinetic energy grid of 751 points between 200 and 275 eV. Unless stated otherwise, a Gaussian FWHM of 1.6 eV was applied. In the Supplementary Material (SM) we give the full details on the dataset and molecular graph generation used in this work. The full set of calculated 6,501 carbon site spectra generated by this work has been made publicly available at https://doi.org/10.5281/zenodo.22285217.

\begin{figure}[!htp]
    \centering
    \includegraphics[width=1.0\columnwidth]{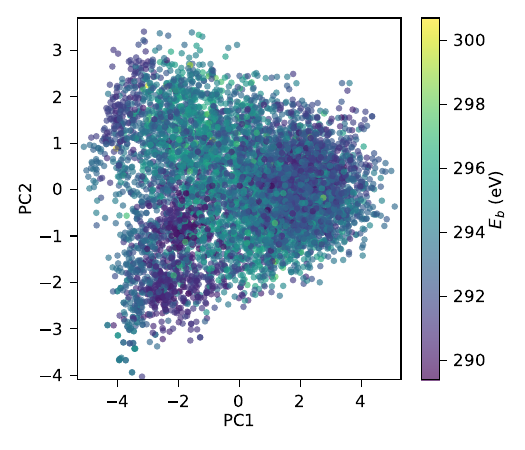}
    \caption{Scatter plot of the two dominant modes from principal component analysis (PCA) of the Auger spectra calculated on the set of QM9 molecules. The color gradient of the scatter points is based on the calculated carbon 1$s$ binding energies.}
    \label{fig:pca}
\end{figure}
 Figure \ref{fig:data} a) shows the train and validation split of our Auger dataset, and details the evaluation data used for AES predictions. Initially a randomly selected sample of 50 molecules was extracted for the hold-out test set, then the remaining molecules underwent a 10-fold cross validation (CV) using the Butina splitting approach.\cite{butina_unsupervised_1999} In addition to the hold-out test, we provide an additional evaluation test set, which contains experimental and calculated spectra for 16 of the 20 organic molecules used in the study benchmarking the electronic structure calculations.\cite{doi:10.1021/acs.jpca.5c01789} We refer to each of these datasets as eval-calc and eval-exp. The structures of the molecules are provided in the SM. The counts of the molecular sizes and atom types for the calculated QM9 (grey) and evaluation (orange) Auger spectra are shown in Figure \ref{fig:data} b). The calculated data used for the train, valid and hold-out sets is imbalanced towards containing more 12-atom molecules and fewer fluorinated molecules, which is inherited from the distribution in the full QM9 database. The evaluation data consists mostly of molecules with fewer than 12 atoms, and therefore examines the model's performance on smaller molecules.
 
\begin{figure*}[!htbp]
    \centering
    \includegraphics[width=1.0\textwidth]{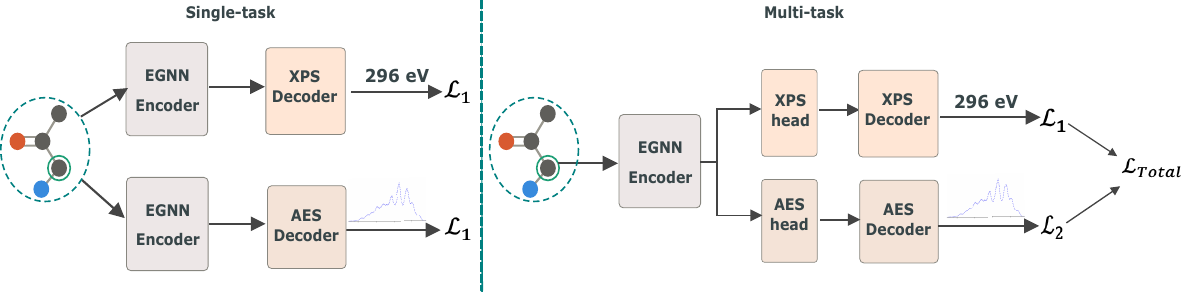}
    \caption{Left: single-task architectures for training separate models to predict carbon 1$s$ $E_{b}$ values in XPS and $f(E_{k})$ in AES. Right: Multi-task architecture for training a single model that predicts both $E_{b}$ and $f(E_{k})$, with a shared graph embedding generated by the EGNN encoder, and separate output heads for each observable. The loss functions for each task ($\mathcal{L}_{1}$ and $\mathcal{L}_{2}$) are combined to the total loss ($\mathcal{L}_{Total}$) via uncertainty weighting in this work.\cite{Kendall_2018_CVPR}}
    \label{fig:gnn}
\end{figure*}
A key difference between the present EGNN architecture and the one previously used to predict carbon 1$s$ $E_{b}$ values,\cite{fouda2026experimentallyaccurategraphneural} is that the dataset now contains both the Auger spectrum ($f(E_{k})$) and the corresponding $E_{b}$ values. This presents an opportunity to implement a multi-task training scheme to develop a model that predicts both $f(E_{k})$ and $E_{b}$ observables from the same weights. This utilizes the inter-task relationship to facilitate inductive knowledge transfer during the training, and enable more efficient model training while potentially improving the single task model prediction accuracy.\cite{Kendall_2018_CVPR,doi:10.1021/acs.jcim.1c00646} Figure \ref{fig:pca} shows the relationship between $f(E_{k})$ and $E_{b}$ in our calculated QM9 subset via principal component analysis of the $f(E_{k})$, with a color gradient corresponding to $E_{b}$. While the variance in $f(E_{k})$ will be influenced by other factors such as the size of the molecule, the analysis shows a relationship by the presence of $E_{b}$ clustering in the two dominant variation modes (PC1 and PC2) of $f(E_{k})$, which can be exploited by multi-task learning.

While multi-task learning is commonplace in computer vision,\cite{Li2024,Zhang_2023_CVPR} natural language processing,\cite{10.1145/1390156.1390177} and speech recognition,\cite{6639081} its application to molecular property,\cite{https://doi.org/10.1002/advs.202412987,10.1063/5.0201681} and particularly spectroscopy prediction, is sparse. The DetaNet\cite{Zou2023} and Molspectra\cite{Liu2026} models apply common architectures to predicting multiple spectroscopic techniques, including infrared, UV-visible, nuclear magnetic resonance and mass spectrometry, but as separate single-task models. Han \textit{et al.}\ applied multi-task training schemes to predicting molecular properties in the QM9 database. They found that architectures containing separate modules for learning a common embedding representation and learning task specific output heads, improve predictions on tasks with complex relationships,\cite{10.1063/1674-0068/cjcp2203055} and outperform previous approaches simply extending a single-task model's output dimension from 1 to number of predicted tasks.\cite{doi:10.1021/acs.jcim.1c00646}

We have applied the hard parameter sharing adaptation to the EGNN model previously used for $E_{b}$ predictions,\cite{fouda_2026_19689244} and a schematic for this is given in Figure \ref{fig:gnn}. The single-task models on the left show how the EGNN encodes the molecular structure into a graph embedding, which is used as an input to either an XPS decoder for $E_{b}$ or an AES decoder for $f(E_{k})$. In the multi-task scheme on the right, the graph embedding is simultaneously passed to two output heads for XPS and AES, which are multi-layer perceptrons (MLPs) generating inputs for the XPS and AES decoders. Each output head contains two fully connected layers followed by the SiLU activation function. In the single and multi-task models, the XPS decoder is a linear projection of the embedding dimension to a dimension of 1 for the scalar $E_{b}$ output. The AES decoder is an MLP with an architecture inspired by the ML XAS literature.\cite{kharel_omnixas_2025,gleason_cuxasnet_2025} First, the node embedding is linearly projected to a higher dimension and passed through layer normalization and Softplus activation, to a second fully connected layer and Softplus activation. A final linear projection to the energy grid dimension, fixed to 751 in this work, is then followed by Softplus activation to re-enforce positive output intensities. The EGNN encoder uses 3 message passing layers with an embedding dimension of 64. In the SM we provide the full set of computational details for the model architecture and training used in this work.
\begin{figure*}[!htbp]
    \centering
    \includegraphics[width=1.0\textwidth]{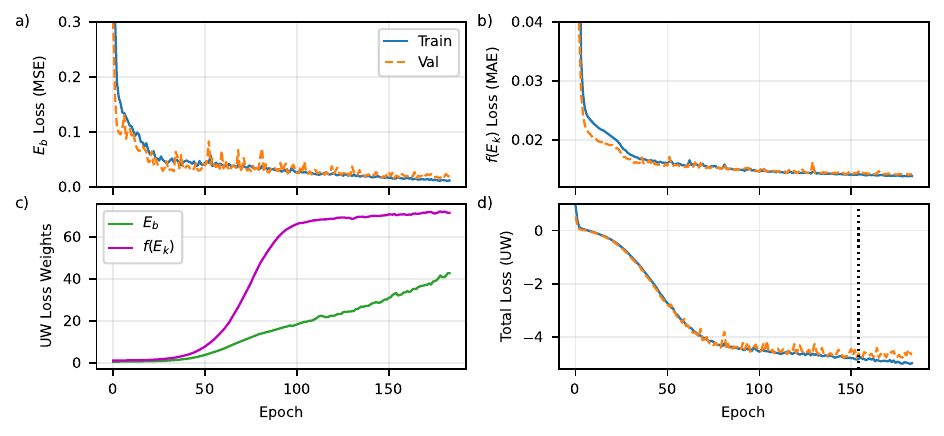}
    \caption{Training curves for the multi-task learning with the MSE loss function for $E_{b}$ prediction and the MAE loss function for $f(E_{k})$ prediction. a) and b) show the training and validation losses per epoch for $E_{b}$ and $f(E_{k})$ respectively. c) the uncertainty loss weights learned during training for each task. The loss weights in c) are applied to the curves in a) and b) and summed to form the total uncertainty weight loss curves in d). The black dotted line in d) indicates the best total validation loss within a patience of 30 epochs, used for the model selection.}
    \label{fig:loss}
\end{figure*}

In order to simultaneously train the model on multiple tasks, the loss functions for each task must be combined for the steepest gradient descent optimization. This work combines the task specific losses via the uncertainty weighted (UW) loss procedure.\cite{Kendall_2018_CVPR} UW loss circumvents manually tuning the weights for each task loss, by instead considering the homoscedastic, or task-dependent, uncertainty. This is a type of model uncertainty that is independent of the uncertainties attributed to the training data size and the nature of the data itself (i.e. the electronic structure theory and graph representation), but depends on the representation of the task and its unit of measure. If we define the prediction of $E_{b}$ as Task 1 and the predictions of $f(E_{k})$ as Task 2 and both tasks use a mean squared error (MSE) loss function ($\mathcal{L}^{MSE}_{1}$ and $\mathcal{L}^{MSE}_{2}$), then the total uncertainty weighted loss ($\mathcal{L}_{Total}$) can then be expressed as,
\begin{equation}\label{eq:uwmse}
\mathcal{L}_{Total} = \frac{1}{2\sigma_{1}^{2}}\mathcal{L}^{MSE}_{1} + \frac{1}{2\sigma_{2}^{2}}\mathcal{L}^{MSE}_{2} + \log\sigma_{1}\sigma_{2},
\end{equation}
where $\sigma$ is the noise parameter for the Gaussian likelihood resulting from the respective MSE loss function. Equation \ref{eq:uwmse} is implemented by initializing $s =\log(\sigma^{2})$ to zero and passing $s$ through the backpropagation procedure that minimizes the model weights, which is more numerically stable than using $\sigma^{2}$.\cite{Kendall_2018_CVPR} As the noise parameter for the task decreases, the task's loss function weights increase. The $\log\sigma_{1}\sigma_{2}$ term in Equation \ref{eq:uwmse} acts as a regularizer that prevents the noise parameter from increasing and collapsing the task loss weight to zero. We refer the reader to the original UW publication by Kendall \textit{et al.}\cite{Kendall_2018_CVPR} for the full derivation. The MSE loss function may not be preferable for all regression tasks and for cases where Task 1 uses $\mathcal{L}^{MSE}_{1}$ and Task 2 uses a mean absolute error (MAE) loss function ($\mathcal{L}^{MAE}_{2}$), the total uncertainty weighted loss is expressed as,
\begin{equation}\label{eq:uwmix}
\mathcal{L}_{Total} = \frac{1}{2\sigma_{1}^{2}}\mathcal{L}^{MSE}_{1} + \frac{1}{b_{2}}\mathcal{L}^{MAE}_{2} + \log\sigma_{1} + \log b_{2},
\end{equation}
where $b_{2}$ is now the noise parameter corresponding to the Laplace likelihood,\cite{Kirchdorfer2025} and $s =\log b_{2}$ and initialized to zero in the training.

Figure \ref{fig:loss} illustrates the UW loss mechanism during the training with the $\mathcal{L}_{1}^{MSE}$ for the $E_{b}$ prediction task and $\mathcal{L}_{2}^{MAE}$ for the $f(E_{k})$ prediction task. The train (blue) and validation (orange) curves for $E_{b}$ and $f(E_{k})$ are given in a) and b) respectively. Between 0 and 100 epochs, the $f(E_{k})$ loss reaches a lower value than the $E_{b}$ loss and therefore its noise parameter is smaller, and the loss weight in c) (green for $E_{b}$ and magenta for $f(E_{k})$) reaches a higher value. This balances the contributions of each task to the total UW loss shown in d). The total loss becomes negative as $\sigma$ for each task decreases, due to the presence of the regularizer terms. The dataset and code for the models used in this work are publicly available at https://doi.org/10.5281/zenodo.22285217 and https://doi.org/10.5281/zenodo.22283453 respectively. The Git repository for the software can be found at https://github.com/afouda11/AugerNet.

\section{Results and Discussion}

\subsection{Overall Model Performance}\label{sec:results}

\begin{table*}[t]
\centering
\renewcommand{\arraystretch}{1.3}
\caption{Summary of the single-task and multi-task EGNN models for predicting the $E_{b}$ and $f(E_{k})$, with respect to the MSE and MAE loss function applied to each task ($\mathcal{L}^{MSE}$ and $\mathcal{L}^{MAE}$). All values are the mean $\pm$ standard deviation over the 10 CV folds. The $E_{b}$ prediction performance metric is the mean absolute error (MAE) in eV for the 113 molecules with experimental carbon 1$s$ binding energies, taken from a previous study.\cite{fouda2026experimentallyaccurategraphneural} The full list of molecules and experimental $E_{b}$ values can be found in the SM. Two metrics are used for the performance of $f(E_{k})$ prediction, the median mean squared error (MSE), reported in units of $10^{-3}$, and the average Pearson's correlation coefficient (PCC). These metrics are given for the hold-out, eval-calc and eval-exp test sets detailed in Figure \ref{fig:data}.}
\label{tab:main}
\begin{tabular*}{\textwidth}{@{\extracolsep{\fill}}ll|ccc|ccc@{}}
\toprule
 Model: Loss Functions & $E_{b}$ Exp. & \multicolumn{3}{c|}{$f(E_{k})$ Median MSE ($\times 10^{-3}$)} & \multicolumn{3}{c}{$f(E_{k})$ Avg. PCC}\\
 & MAE (eV) & Hold-out & Eval-calc & Eval-exp & Hold-out & Eval-calc & Eval-exp\\
\midrule
Single-task: $\mathcal{L}^{MSE}$ & 0.32 $\pm$ 0.02 & 6.0 $\pm$ 0.5 & 23.3 $\pm$ 3.1 & 42.1 $\pm$ 5.3 & 0.96 $\pm$ 0.00 & 0.81 $\pm$ 0.01 & 0.85 $\pm$ 0.01\\
Single-task: $\mathcal{L}^{MAE}$ & 0.34 $\pm$ 0.02 & 5.0 $\pm$ 0.2 & 20.7 $\pm$ 2.8 & 48.7 $\pm$ 3.9 & 0.96 $\pm$ 0.00 & 0.80 $\pm$ 0.02 & 0.83 $\pm$ 0.01\\
Multi-task:  $\big(\mathcal{L}_1^{MSE},\mathcal{L}_2^{MSE}\big)$ & 0.32 $\pm$ 0.03 & 7.1 $\pm$ 0.8 & 23.6 $\pm$ 2.8 & 37.0 $\pm$ 3.7 & 0.96 $\pm$ 0.00 & 0.83 $\pm$ 0.01 & 0.87 $\pm$ 0.01\\
Multi-task:  $\big(\mathcal{L}_1^{MAE},\mathcal{L}_2^{MAE}\big)$ & 0.32 $\pm$ 0.02 & 6.4 $\pm$ 0.4 & 21.1 $\pm$ 1.3 & 38.6 $\pm$ 3.4 & 0.96 $\pm$ 0.00 & 0.83 $\pm$ 0.01 & 0.87 $\pm$ 0.01\\
Multi-task:  $\big(\mathcal{L}_1^{MSE},\mathcal{L}_2^{MAE}\big)$ & 0.32 $\pm$ 0.01 & 6.3 $\pm$ 0.5 & 21.6 $\pm$ 1.6 & 37.0 $\pm$ 3.4 & 0.96 $\pm$ 0.00 & 0.83 $\pm$ 0.01 & 0.87 $\pm$ 0.01\\
Multi-task:  $\big(\mathcal{L}_1^{MAE},\mathcal{L}_2^{MSE}\big)$ & 0.33 $\pm$ 0.03 & 7.3 $\pm$ 0.6 & 23.9 $\pm$ 2.1 & 36.2 $\pm$ 3.3 & 0.96 $\pm$ 0.00 & 0.82 $\pm$ 0.01 & 0.87 $\pm$ 0.01\\
\bottomrule
\end{tabular*}
\vspace{2pt}
\footnotesize
For the calculated Eval. spectra against the experimental Eval. spectra the Avg. PCC is 0.78 and the Median MSE is 67.2 $\times 10^{-3}$. (Model-independent; identical for every row.)
\end{table*}

The results of the single-task and multi-task EGNN models with respect to both $\mathcal{L}^{MSE}$ and $\mathcal{L}^{MAE}$ are presented as the mean and standard deviations across the 10 CV folds in Table \ref{tab:main}. For $E_{b}$ prediction, the models are evaluated with the MAE against a set of 113 molecules with experimental carbon 1$s$ $E_{b}$ values, previously used to evaluate single-task EGNN $E_{b}$ predictions,\cite{fouda2026experimentallyaccurategraphneural} we refer the reader to this publication and the present SM for the full list of molecules and experimental $E_{b}$ values. To evaluate the AES predictions, two complementary metrics are provided, the median (MSE) and the average Pearson's correlation coefficient (PCC). Taking the median of the spectral MSE was previously found to be robust to the differences between model error distributions in the prediction of XAS spectra.\cite{kharel_omnixas_2025} The PCC is a measure of the lineshape similarity, based on the cosine similarity between two vectors and their means. Previously, this metric was found to be robust to spectral noise and peak broadening in the evaluation of XAS spectra.\cite{Suzuki2019} 

Across the 10 CV folds, the single- and multi-task models show similar overall predictive performance for both tasks, loss functions, evaluation metrics and datasets. This indicates that the physical connection, and thus correlation, between $E_{b}$ and $f(E_{k})$ is sufficient for both tasks to share a common graph embedding representation. For reference, we have included the performance metrics for the eval-exp dataset and note that the training data level of theory has a median MSE of 0.067 and average PCC of 0.78 against the experimental spectra. While all models give better agreement to experiment than the calculated spectra, this is not a reliable metric for the model performance; Subsection \ref{sec:discuss} discusses how the predicted spectra tends to be broader, with less sharp peaks than the calculated spectra. This happens to show better agreement to experiment at the Gaussian FWHM of 1.6 eV.
\begin{figure*}[!htbp]
    \centering
    \includegraphics[width=1.0\textwidth]{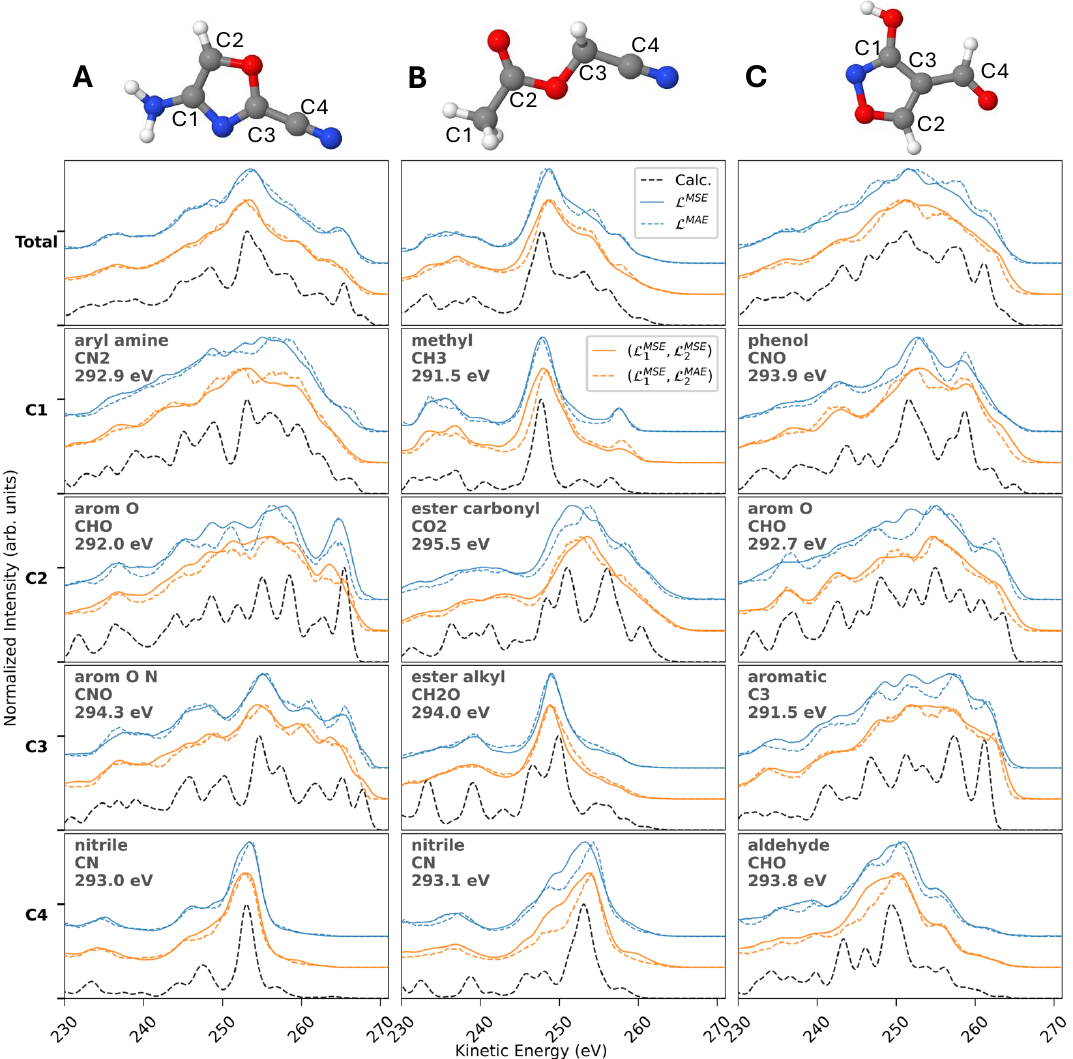}
    \caption{Auger spectra for 3 hold-out test set molecules labeled A, B and C, each molecule has 4 unique carbon environments. The top row is the combination of all four carbon 1$s$ spectra and the panels below correspond to the carbons labeled in the structures above. The single-task models with the $\mathcal{L}^{MSE}$ (solid blue) and $\mathcal{L}^{MAE}$ (dashed blue), and multi-task models with $\big(\mathcal{L}_1^{MSE},\mathcal{L}_2^{MSE}\big)$ (solid orange) and $\big(\mathcal{L}_1^{MSE},\mathcal{L}_2^{MAE}\big)$ (dashed orange) spectra are compared to the calculated (black dashed) spectra.}
    \label{fig:holdout}
\end{figure*}

For $E_{b}$ prediction, the range of the mean MAEs across all the models is 0.02 eV, the single-task $\mathcal{L}^{MAE}$ model gives the worse overall performance and the multi-task $\big(\mathcal{L}_1^{MSE},\mathcal{L}_2^{MAE}\big)$ model has the best overall performance by a small margin, with the same mean MAE as the single-task $\mathcal{L}^{MSE}$ and multi-task $\big(\mathcal{L}_1^{MAE},\mathcal{L}_2^{MAE}\big)$ models of 0.32 eV and a slightly improved standard deviation of 0.01 eV. The overall $E_{b}$ results also indicate a multi-task advantage with respect to the choice of the loss function, as both multi-task models using $\mathcal{L}_{1}^{MAE}$, have an improved performance over the single task $\mathcal{L}^{MAE}$ model.
\begin{figure*}[!htbp]
    \centering
    \includegraphics[width=1.0\textwidth]{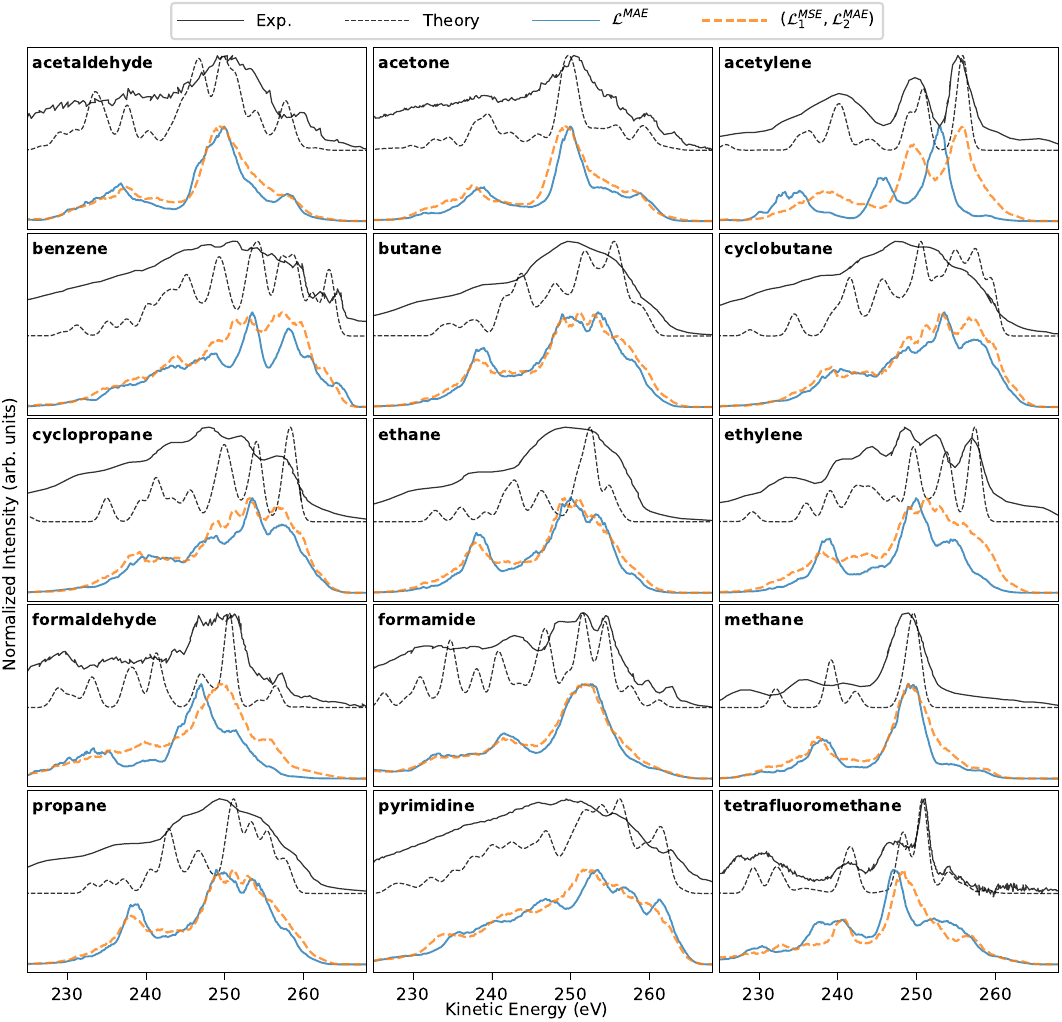}
    \caption{Auger spectra for evaluation test set of molecules, with individual carbon spectra combined to the molecular spectra. A single-task model with $\mathcal{L}^{MAE}$ (solid blue) and multi-task models with $\big(\mathcal{L}_1^{MSE},\mathcal{L}_2^{MAE}\big)$ (dashed orange) spectra are compared to the calculated (eval-calc) (black dashed) and experimental (eval-exp) (black solid) spectra. The experimental spectra for each molecule are taken from the following references: acetaldehyde,\cite{10.1063/1.461687} acetone,\cite{10.1063/1.461687} acetylene,\cite{10.1063/1.440831} benzene,\cite{10.1063/1.1290029} butane,\cite{10.1063/1.440015} cyclobutane,\cite{10.1063/1.440831} cyclopropane,\cite{10.1063/1.440831} ethane,\cite{10.1063/1.440015} ethylene,\cite{10.1063/1.440831} formaldehyde,\cite{10.1063/1.461687} formamide,\cite{10.1063/1.461687} methane,\cite{10.1063/1.440015} propane,\cite{10.1063/1.440015} pyrimidine\cite{10.1063/1.2993317} and tetrafluoromethane.\cite{MNeeb_1997}. Predictions on oxazole\cite{10.1063/5.0122088} are included in Table \ref{tab:main}'s metrics, but not in the present figure.}
    \label{fig:eval}
\end{figure*}

For $f(E_{k})$ prediction, all models give the same performance for the average PCC values on the hold-out set with a CV fold-averaged average PCC value of 0.96. The hold-out median MSE values however, have a fold averaged range of $2.3\times10^{-3}$. This indicates that the median MSE is a more precise metric for evaluating the model performance. For the eval-calc dataset, the average PCC values show a range of 0.03 across the fold averaged values, and the median MSE range in fold averaged values is $3.2\times10^{-3}$. The larger deviations in performance on the eval-calc dataset are expected due to the smaller number of molecules in this dataset, which have a different molecular size and atom type distribution to the dataset the hold-out molecules were sampled from (see Figure \ref{fig:data}).

In all cases, the median MSE and average PCC values in the hold-out and eval-calc datasets are improved when $\mathcal{L}^{MAE}$ is used instead of $\mathcal{L}^{MSE}$ for the $f(E_{k})$ prediction. $\mathcal{L}^{MAE}$ is better at describing finer peak structures by applying a less strict penalty to deviations from the target line-shape during the training. Unlike the  $E_{b}$ prediction task, there is no apparent multi-task advantage with respect to stability of the model performance and the choice of the loss function. The median MSE values of the hold-out and eval-calc datasets show that the multi-task training does not appear to correct for the reduced performance of $\mathcal{L}^{MSE}$ in the single-task models. The multi-task $\big(\mathcal{L}_1^{MAE},\mathcal{L}_2^{MSE}\big)$ and $\big(\mathcal{L}_1^{MSE},\mathcal{L}_2^{MSE}\big)$ perform worse than the both single-task models. However, the performance of the $f(E_{k})$ prediction appears to be relatively stable with respect to the choice of the loss function for the $E_b$ task as the fold averaged hold-out median MSE values for the $\big(\mathcal{L}_1^{MSE},\mathcal{L}_2^{MSE}\big)$ and $\big(\mathcal{L}_1^{MAE},\mathcal{L}_2^{MSE}\big)$ multi-task models are 7.1 and 7.3 respectively, and for $\big(\mathcal{L}_1^{MAE},\mathcal{L}_2^{MAE}\big)$ and $\big(\mathcal{L}_1^{MSE},\mathcal{L}_2^{MAE}\big)$ they are 6.4 and 6.3 respectively. This insensitivity to the choice of $E_b$ task task loss is also reflected in the fold averaged eval-calc values.

The lack of a clear multi-task advantage for the $f(E_{k})$ prediction possibly relates to the increased complexity of this task. Despite this, the overall agreement of the CV fold averaged results between the multi-task and single-task models is good within the context of the available evaluation datasets. In the following discussion, to give a fair comparison between the models, the results are taken from a single fold (fold 8), This fold was selected for its validation loss being close to the fold averaged validation loss across all models, and therefore not particularly advantageous for any of the models shown in the following results.

\begin{figure}[!htbp]
    \centering
    \includegraphics[width=1.0\columnwidth]{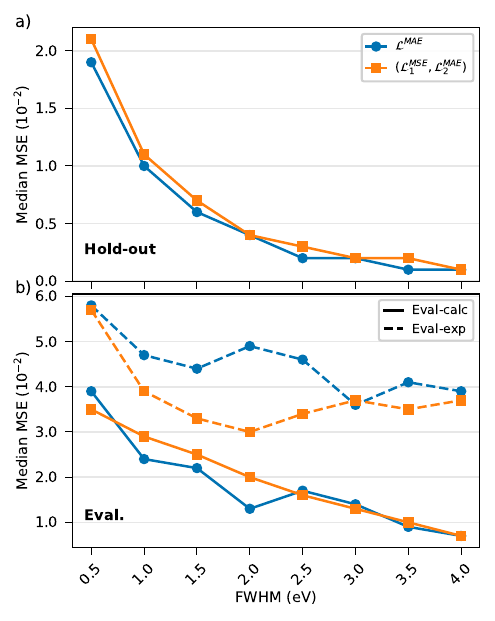}
    \caption{Plot showing the effect of the Gaussian FWHM used to fit the spectrum on the median MSE of the a) hold-out and b) evaluation (eval-calc: solid, eval-exp: dashed) test sets. The single-task $\mathcal{L}^{MAE}$ model (blue) is compared to the multi-task $\big(\mathcal{L}_1^{MSE},\mathcal{L}_2^{MAE}\big)$ (orange) model.}
    \label{fig:fwhm}
\end{figure}

\subsection{What Do the Models Learn About AES?}\label{sec:discuss}

Figure \ref{fig:holdout} examines the predicted and calculated Auger spectra for 3 molecules from the hold-out dataset labeled \textbf{A}, \textbf{B} and \textbf{C}, which are shown above the spectra. These molecules are showcased as they each contain 4 unique carbon environments. The combined molecular spectra are given in the top row, and the individual carbon spectra are presented in the rows below. Each carbon is labeled C1-C4, which correspond to the labels in the structures above the spectra. The environment class and calculated $E_b$ for each carbon are provided in the panels. The calculated spectra (black dashed lines) are compared to the single task models with the $\mathcal{L}^{MSE}$ (blue solid) and $\mathcal{L}^{MAE}$ (blue dashed) loss functions and the $\big(\mathcal{L}_1^{MSE},\mathcal{L}_2^{MSE}\big)$ (orange solid) and $\big(\mathcal{L}_1^{MSE},\mathcal{L}_2^{MAE}\big)$ (orange dashed) multi-task models. These multi-task models were selected to show the effect of switching from $\mathcal{L}_2^{MSE}$ to $\mathcal{L}_2^{MAE}$ on the line-shape prediction. Generally, both single-task and multi-task models give a good predictive accuracy of the calculated spectra, but as mentioned earlier the predicted spectra tend to be more broad with less well defined peak structures. Furthermore, this broadness is generally increased for the multi-task predictions. This is evidenced by the aromatic carbons at molecule \textbf{A} C2 and C3, and molecule \textbf{C}  C2 and C3, where the single-task models resolve a finer peak structure than the multi-task models. In some cases, such as C2 \textbf{A}, C4 \textbf{B} and C1 \textbf{C}, switching from $\mathcal{L}_2^{MSE}$ to $\mathcal{L}_2^{MAE}$ in the multi-task model appears to recover some of the finer peak structures. However, in most cases presented, switching to the less strict $\mathcal{L}_2^{MAE}$ loss function in the multi-task models shows little advantage to using to stricter $\mathcal{L}^{MSE}$ loss in the single-task model.

The single-task and multi-task models describe common lineshape features between common carbon environments across different molecules. For the terminal nitrile carbons C4 in molecules \textbf{A} and \textbf{B}, all models capture the single sharp peak between 250 and 260 eV which dominates the lineshape, and for the aromatic carbons in molecules \textbf{A} and \textbf{C}, all models capture the significant intensity spread spanning the full kinetic energy range. This demonstrates the potential of EGNN models, in both the single and multi task settings to provide a rapid, approximate analysis of AES lineshapes and make carbon specific assignments in high-resolution AES measurements. 

In Figure \ref{fig:eval}, the model performance on the eval-calc (black dashed) and eval-exp (black solid) spectra supports this further, which compares the single-task $\mathcal{L}^{MAE}$ model (blue solid) with the multi-task $\big(\mathcal{L}_1^{MSE},\mathcal{L}_2^{MAE}\big)$ model (orange dashed). The combined molecular spectra are shown, but acetylene, benzene, cyclobutane, cyclopropane, ethane, ethylene, formaldehyde, methane and tetrafluoromethane all contain a single carbon environment and only acetaldehyde, acetone, butane, propane and pyrimidine contain a combination of multiple carbon spectra. For consistency with the previous study which benchmarked the theory,\cite{doi:10.1021/acs.jpca.5c01789} the calculated and thus predicted spectra have all been shifted by -2 eV to align them better to the experiment. While for most cases this shift aligns the highest energy peaks of predictions and calculations to experiments well. In some cases this shift causes a increased misalignment of the predicted spectra to lower kinetic energies with respect to the calculations and experiment. For example in tetrafluoromethane, both the single and multi-task models major peak appears at a lower kinetic energy to the calculated and experimental major peak about 250 eV. However, in other cases such as acetylene, ethylene and formaledyde the multi-task model has a better alignment of spectrum kinetic energies than the single-task model to the calculations and experiment. However, acetylene is the only case where the lineshape of the multi-task predictions shows an improvement over the single-task prediction. Generally, the models give a good reproduction of the overall lineshape with respect to the calculations and the experiment, though the performance of both single-task and multi-task models is poor for some of the smaller molecules such as ethane, formaldehyde and formamide, which is not surprising considering the size imbalances in the training data (see Figure \ref{fig:data}).
\begin{figure*}[!htbp]
    \centering
    \includegraphics[width=1.0\textwidth]{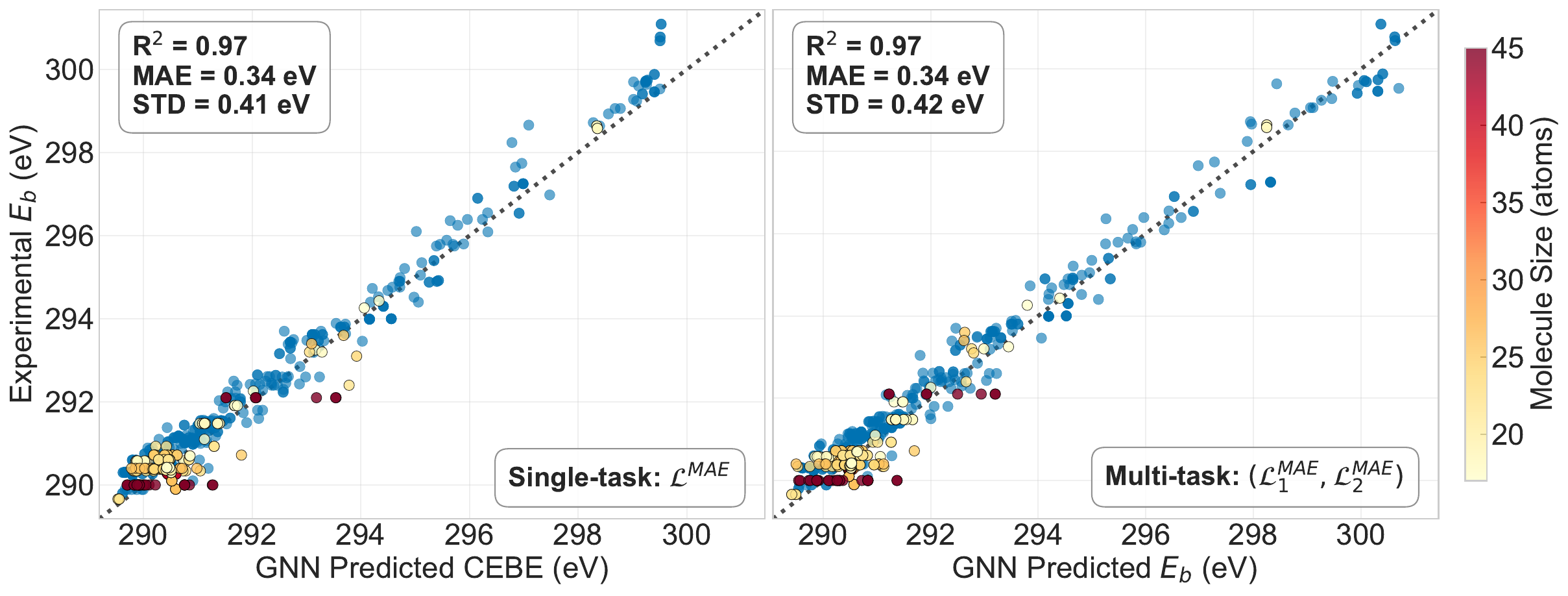}
    \caption{Scatter plots for the single-task $\mathcal{L}^{MAE}$ model (left) and multi-task $\big(\mathcal{L}_1^{MAE},\mathcal{L}_2^{MAE}\big)$ (right) model predictions of experimental carbon 1$s$ $E_{b}$ values on the 113 organic molecules previously used to evaluate EGNN predictions.\cite{fouda2026experimentallyaccurategraphneural} Molecules with under 16 atoms are shown by blue scatter points and molecules with 16 atoms or more are shown by the yellow-to-red color gradient.}
    \label{fig:cebe}
\end{figure*}

The choice to train the EGNN models on spectra fitted with a 1.6 eV FWHM Gaussian was somewhat arbitrary, but tests how well the models can resolve complex peak structures. It is clear however, that the available experimental reference spectra are much broader and a FWHM of 3.8 eV was previously used to benchmark the theory to experiment.\cite{doi:10.1021/acs.jpca.5c01789} Figure \ref{fig:fwhm}, shows the model's accuracy with respect to the FWHM over a 0.5 to 4.0 eV range. The median MSE of the single-task $\mathcal{L}^{MAE}$ (blue solid) and multi-task $\big(\mathcal{L}_1^{MSE},\mathcal{L}_2^{MAE}\big)$ (orange solid) models is shown with respect to the hold-out set in a). b) shows the performance on the eval-calc (solid) and eval-exp (dashed) datasets. As expected, the single-task and multi-task models are more accurate at reproducing the calculated spectra at broader FWHM, and show a similar on the hold-out set. On the eval-calc dataset the single-task model shows an improved performance below a FWHM of 2.5 eV. This dataset contains a larger proportion of smaller molecules, which have spectra with sharper peak structures and thus better predicted by the single-task model. The experimental spectra are kept fixed at their reported widths and the fluctuations in model performance across the FWHM range  reflects the diversity molecules within this smaller dataset.

Finally, we show a scatter plot of the EGNN predictions of the 113 molecules with experimental carbon 1$s$ $E_{b}$ values used for the MAE values in Table \ref{tab:main}. The left panel shows the single-task $\mathcal{L}^{MAE}$ model and the right panel has the multi-task $\big(\mathcal{L}_1^{MAE},\mathcal{L}_2^{MAE}\big)$ model. The yellow to red color gradient indicates binding energies from molecules with more than 16 atoms, to show the size transferability of the model to larger organic molecules. Whilst the R$^2$, MAE and standard deviation (STD) performance metrics have similar values for the single and multi-task models, a minor multi-task advantage is observed for the two highest binding energy carbons with experimental values above 300 eV. In this region, the single-task model reaches an upper limit to the energies it can predict, as shown by the vertical scatter points around 299 eV. This energy region contains highly fluorinated carbons, which are underrepresented in the training data (see Figure \ref{fig:data}), and despite this the multi-task model is able to more accurately predict these as values above 300 eV. This indicates that a shared embedding learned across multiple related tasks can improve the generalization of predictive models for data-points from low population regions in the training data.  

\section{Conclusion}

To conclude, by constructing a dataset of AES spectra with efficient multiconfigurational chemistry calculations, we have demonstrated that multi-task training of the observables in XPS and AES is possible with EGNN models. The predicted spectra, in both the single and multi task regimes give good reproductions of calculated and experimental lineshapes, and experimental $E_{b}$ values. Multi-task training schemes for predicting spectroscopy observables open the possibility of building universal x-ray spectroscopy models trained on multi-modal datasets. Inductive knowledge transfer and shared embedding spaces are well suited to x-ray spectroscopy, as the different techniques share a common relationship to the local bond environment about the atom which absorbs the x-ray.

In order to overcome the model's dependency on a fixed FWHM, future models could train over spectra fitted across a range of FWHM values and the FWHM could be included as a graph or node level input feature. A more sophisticated approach however, would be to utilize multi-task training to develop a model that predicts the Auger stick spectrum, then apply the FWHM as a postprocessing step, in the same way that calculated spectra are presented. Such a model would require separate decoders for the energies and intensities for both the singlet and triplet final states, yielding a 4 task model solely for AES prediction. This lies beyond the scope of the current work and would require a larger and more diverse training data sample.

The complexity of AES limits the availability of its training and evaluation data, and thus current scope of ML applications in this field. We hope that this work, and the continued development of AES calculations and experimental measurements can inspire more data-driven approaches exploiting the wealth of information encoded in the Auger-spectrum lineshape and its sensitivity to the local bond environment.

\begin{acknowledgments}
We would like to thank Linda Young for her support and recommendations on the manuscript. A.E.A.F. is grateful for the support from the Eric and Wendy Schmidt AI in Science Postdoctoral Fellowship, a Schmidt Futures Program. This work was supported by the U.S. Department of Energy, Office of Basic Energy Sciences, Division of Chemical Sciences, Geosciences, and Biosciences at Argonne National Laboratory, under contract DE-AC02-06CH11357.
\end{acknowledgments}

\section*{Data and Software Availability Statement}
The data and the software are freely available at https://doi.org/10.5281/zenodo.22285217 and https://doi.org/10.5281/zenodo.22283453 respectively. The Git repository for the software can be found at https://github.com/afouda11/AugerNet.

\section*{Supplementary Material}
The Supplementary Material (SM) for this work includes the full set of computational details for the database, molecular graphs, model architecture and training parameters. Furthermore the SM contains a table of the all the molecules and experimental carbon 1$s$ binding energies used to evaluate the models in this work. 
\section*{References}
\bibliography{literature}

@article{porcelli_photoemission_2025,
	title = {Photoemission spectroscopy of organic molecules using plane wave/pseudopotential density functional theory and machine learning: {A} comprehensive and predictive computational protocol for isolated molecules, molecular aggregates, and organic thin films},
	volume = {162},
	issn = {0021-9606},
	shorttitle = {Photoemission spectroscopy of organic molecules using plane wave/pseudopotential density functional theory and machine learning},
	url = {https://doi.org/10.1063/5.0272583},
	doi = {10.1063/5.0272583},
	number = {24},
	urldate = {2025-11-25},
	journal = {The Journal of Chemical Physics},
	author = {Porcelli, Francesco and Filippone, Francesco and Colasante, Emanuela and Mattioli, Giuseppe},
	month = jun,
	year = {2025},
	pages = {244101},
}

@misc{satorras_en_2022,
	title = {E(n) {Equivariant} {Graph} {Neural} {Networks}},
	url = {http://arxiv.org/abs/2102.09844},
	doi = {10.48550/arXiv.2102.09844},
	language = {en},
	urldate = {2026-02-25},
	publisher = {arXiv},
	author = {Satorras, Victor Garcia and Hoogeboom, Emiel and Welling, Max},
	month = feb,
	year = {2022},
	note = {arXiv:2102.09844 [cs]},
}

@article{rankine_progress_2021,
	title = {Progress in the {Theory} of {X}-ray {Spectroscopy}: {From} {Quantum} {Chemistry} to {Machine} {Learning} and {Ultrafast} {Dynamics}},
	volume = {125},
	copyright = {https://doi.org/10.15223/policy-029},
	issn = {1089-5639, 1520-5215},
	shorttitle = {Progress in the {Theory} of {X}-ray {Spectroscopy}},
	url = {https://pubs.acs.org/doi/10.1021/acs.jpca.0c11267},
	doi = {10.1021/acs.jpca.0c11267},
	language = {en},
	number = {20},
	urldate = {2026-03-04},
	journal = {The Journal of Physical Chemistry A},
	author = {Rankine, C. D. and Penfold, T. J.},
	month = may,
	year = {2021},
	pages = {4276--4293},
}

@article{golze_accurate_2022,
	title = {Accurate {Computational} {Prediction} of {Core}-{Electron} {Binding} {Energies} in {Carbon}-{Based} {Materials}: {A} {Machine}-{Learning} {Model} {Combining} {Density}-{Functional} {Theory} and \textit{{GW}}},
	volume = {34},
	copyright = {https://creativecommons.org/licenses/by/4.0/},
	issn = {0897-4756, 1520-5002},
	shorttitle = {Accurate {Computational} {Prediction} of {Core}-{Electron} {Binding} {Energies} in {Carbon}-{Based} {Materials}},
	url = {https://pubs.acs.org/doi/10.1021/acs.chemmater.1c04279},
	doi = {10.1021/acs.chemmater.1c04279},
	language = {en},
	number = {14},
	urldate = {2026-03-04},
	journal = {Chemistry of Materials},
	author = {Golze, Dorothea and Hirvensalo, Markus and Hernández-León, Patricia and Aarva, Anja and Etula, Jarkko and Susi, Toma and Rinke, Patrick and Laurila, Tomi and Caro, Miguel A.},
	month = jul,
	year = {2022},
	pages = {6240--6254},
}

@article{zarrouk_experiment-driven_2024,
	title = {Experiment-{Driven} {Atomistic} {Materials} {Modeling}: {A} {Case} {Study} {Combining} {X}-{Ray} {Photoelectron} {Spectroscopy} and {Machine} {Learning} {Potentials} to {Infer} the {Structure} of {Oxygen}-{Rich} {Amorphous} {Carbon}},
	volume = {146},
	copyright = {https://creativecommons.org/licenses/by/4.0/},
	issn = {0002-7863, 1520-5126},
	shorttitle = {Experiment-{Driven} {Atomistic} {Materials} {Modeling}},
	url = {https://pubs.acs.org/doi/10.1021/jacs.4c01897},
	doi = {10.1021/jacs.4c01897},
	language = {en},
	number = {21},
	urldate = {2026-03-04},
	journal = {Journal of the American Chemical Society},
	author = {Zarrouk, Tigany and Ibragimova, Rina and Bartók, Albert P. and Caro, Miguel A.},
	month = may,
	year = {2024},
	pages = {14645--14659},
}

@article{sun_machine_2022,
	title = {Machine {Learning} {Predicts} the {X}-ray {Photoelectron} {Spectroscopy} of the {Solid} {Electrolyte} {Interface} of {Lithium} {Metal} {Battery}},
	volume = {13},
	copyright = {https://doi.org/10.15223/policy-029},
	issn = {1948-7185, 1948-7185},
	url = {https://pubs.acs.org/doi/10.1021/acs.jpclett.2c02222},
	doi = {10.1021/acs.jpclett.2c02222},
	language = {en},
	number = {34},
	urldate = {2026-03-04},
	journal = {The Journal of Physical Chemistry Letters},
	author = {Sun, Qintao and Xiang, Yan and Liu, Yue and Xu, Liang and Leng, Tianle and Ye, Yifan and Fortunelli, Alessandro and Goddard, William A and Cheng, Tao},
	month = sep,
	year = {2022},
	pages = {8047--8054},
}

@article{tripathy_chemical_2024,
	title = {Chemical space-informed machine learning models for rapid predictions of x-ray photoelectron spectra of organic molecules},
	volume = {5},
	issn = {2632-2153},
	url = {https://iopscience.iop.org/article/10.1088/2632-2153/ad871d},
	doi = {10.1088/2632-2153/ad871d},
	language = {en},
	number = {4},
	urldate = {2026-03-04},
	journal = {Machine Learning: Science and Technology},
	author = {Tripathy, Susmita and Das, Surajit and Jindal, Shweta and Ramakrishnan, Raghunathan},
	month = dec,
	year = {2024},
	pages = {045023},
}

@article{schutt_equivariant_nodate,
	title = {Equivariant {Message} {Passing} for the {Prediction} of {Tensorial} {Properties} and {Molecular} {Spectra}},
	language = {en},
	author = {Schütt, Kristof T and Unke, Oliver T and Gastegger, Michael},
}

@article{buterez_transfer_2024,
	title = {Transfer learning with graph neural networks for improved molecular property prediction in the multi-fidelity setting},
	volume = {15},
	issn = {2041-1723},
	url = {https://www.nature.com/articles/s41467-024-45566-8},
	doi = {10.1038/s41467-024-45566-8},
	language = {en},
	number = {1},
	urldate = {2026-03-04},
	journal = {Nature Communications},
	author = {Buterez, David and Janet, Jon Paul and Kiddle, Steven J. and Oglic, Dino and Lió, Pietro},
	month = feb,
	year = {2024},
	pages = {1517},
}

@article{xu_pretrained_2025,
	title = {Pretrained {E}(3)-equivariant message-passing neural networks with multi-level representations for organic molecule spectra prediction},
	volume = {11},
	issn = {2057-3960},
	url = {https://www.nature.com/articles/s41524-025-01698-z},
	doi = {10.1038/s41524-025-01698-z},
	language = {en},
	number = {1},
	urldate = {2026-03-04},
	journal = {npj Computational Materials},
	author = {Xu, Yuzhi and Bian, Daqian and Ju, Cheng-Wei and Zhao, Fanyu and Xie, Pujun and Wang, Yuanqing and Hu, Wei and Sun, Zhenrong and Zhang, John Z. H. and Zhu, Tong},
	month = jul,
	year = {2025},
	pages = {203},
}

@article{rankine_accurate_2022,
	title = {Accurate, affordable, and generalizable machine learning simulations of transition metal x-ray absorption spectra using the {XANESNET} deep neural network},
	volume = {156},
	issn = {0021-9606, 1089-7690},
	url = {https://pubs.aip.org/jcp/article/156/16/164102/2841051/Accurate-affordable-and-generalizable-machine},
	doi = {10.1063/5.0087255},
	language = {en},
	number = {16},
	urldate = {2026-03-05},
	journal = {The Journal of Chemical Physics},
	author = {Rankine, C. D. and Penfold, T. J.},
	month = apr,
	year = {2022},
	pages = {164102},
}

@article{rankine_deep_2020,
	title = {A {Deep} {Neural} {Network} for the {Rapid} {Prediction} of {X}-ray {Absorption} {Spectra}},
	volume = {124},
	copyright = {https://doi.org/10.15223/policy-029},
	issn = {1089-5639, 1520-5215},
	url = {https://pubs.acs.org/doi/10.1021/acs.jpca.0c03723},
	doi = {10.1021/acs.jpca.0c03723},
	language = {en},
	number = {21},
	urldate = {2026-03-05},
	journal = {The Journal of Physical Chemistry A},
	author = {Rankine, C. D. and Madkhali, M. M. M. and Penfold, T. J.},
	month = may,
	year = {2020},
	pages = {4263--4270},
}

@article{madkhali_role_2020,
	title = {The {Role} of {Structural} {Representation} in the {Performance} of a {Deep} {Neural} {Network} for {X}-ray {Spectroscopy}},
	volume = {25},
	issn = {1420-3049},
	url = {https://www.mdpi.com/1420-3049/25/11/2715},
	doi = {10.3390/molecules25112715},
	language = {en},
	number = {11},
	urldate = {2026-03-05},
	journal = {Molecules},
	author = {Madkhali, Marwah M.M. and Rankine, Conor D. and Penfold, Thomas J.},
	month = jun,
	year = {2020},
	pages = {2715},
}

@article{luder_machine_2025,
	title = {Machine learning approach to predict {L} -edge x-ray absorption spectra of light transition metal ion compounds},
	volume = {111},
	issn = {2469-9950, 2469-9969},
	url = {https://link.aps.org/doi/10.1103/PhysRevB.111.085110},
	doi = {10.1103/PhysRevB.111.085110},
	language = {en},
	number = {8},
	urldate = {2026-03-05},
	journal = {Physical Review B},
	author = {Lüder, Johann},
	month = feb,
	year = {2025},
	pages = {085110},
}

@article{zhan_graph_2025,
	title = {A {Graph} {Neural} {Network}-{Based} {Approach} to {XANES} {Data} {Analysis}},
	volume = {129},
	copyright = {https://doi.org/10.15223/policy-029},
	issn = {1089-5639, 1520-5215},
	url = {https://pubs.acs.org/doi/10.1021/acs.jpca.4c05119},
	doi = {10.1021/acs.jpca.4c05119},
	language = {en},
	number = {4},
	urldate = {2026-03-05},
	journal = {The Journal of Physical Chemistry A},
	author = {Zhan, Fei and Yao, Haodong and Geng, Zhi and Zheng, Lirong and Yu, Can and Han, Xue and Song, Xueqi and Chen, Shuguang and Zhao, Haifeng},
	month = jan,
	year = {2025},
	pages = {874--884},
}

@article{kotobi_integrating_2023-1,
	title = {Integrating {Explainability} into {Graph} {Neural} {Network} {Models} for the {Prediction} of {X}-ray {Absorption} {Spectra}},
	volume = {145},
	copyright = {https://creativecommons.org/licenses/by/4.0/},
	issn = {0002-7863, 1520-5126},
	url = {https://pubs.acs.org/doi/10.1021/jacs.3c07513},
	doi = {10.1021/jacs.3c07513},
	language = {en},
	number = {41},
	urldate = {2026-03-05},
	journal = {Journal of the American Chemical Society},
	author = {Kotobi, Amir and Singh, Kanishka and Höche, Daniel and Bari, Sadia and Meißner, Robert H. and Bande, Annika},
	month = oct,
	year = {2023},
	pages = {22584--22598},
}

@article{carbone_machine-learning_2020,
	title = {Machine-{Learning} {X}-{Ray} {Absorption} {Spectra} to {Quantitative} {Accuracy}},
	volume = {124},
	issn = {0031-9007, 1079-7114},
	url = {https://link.aps.org/doi/10.1103/PhysRevLett.124.156401},
	doi = {10.1103/PhysRevLett.124.156401},
	language = {en},
	number = {15},
	urldate = {2026-03-05},
	journal = {Physical Review Letters},
	author = {Carbone, Matthew R. and Topsakal, Mehmet and Lu, Deyu and Yoo, Shinjae},
	month = apr,
	year = {2020},
	pages = {156401},
}

@article{kharel_omnixas_2025,
	title = {{OmniXAS}: {A} universal deep-learning framework for materials x-ray absorption spectra},
	volume = {9},
	issn = {2475-9953},
	shorttitle = {{OmniXAS}},
	url = {https://link.aps.org/doi/10.1103/PhysRevMaterials.9.043803},
	doi = {10.1103/PhysRevMaterials.9.043803},
	language = {en},
	number = {4},
	urldate = {2026-03-05},
	journal = {Physical Review Materials},
	author = {Kharel, Shubha R. and Meng, Fanchen and Qu, Xiaohui and Carbone, Matthew R. and Lu, Deyu},
	month = apr,
	year = {2025},
	pages = {043803},
}

@article{gleason_cuxasnet_2025,
	title = {{CuXASNet}: {Rapid} and accurate prediction of copper {L} -edge x-ray absorption spectra using machine learning},
	volume = {9},
	issn = {2475-9953},
	shorttitle = {{CuXASNet}},
	url = {https://link.aps.org/doi/10.1103/xz7f-srky},
	doi = {10.1103/xz7f-srky},
	language = {en},
	number = {7},
	urldate = {2026-03-05},
	journal = {Physical Review Materials},
	author = {Gleason, Samuel P. and Carbone, Matthew R. and Lu, Deyu and Ciston, Jim},
	month = jul,
	year = {2025},
	pages = {073803},
}

@article{bartok_representing_2013,
	title = {On representing chemical environments},
	volume = {87},
	copyright = {http://link.aps.org/licenses/aps-default-license},
	issn = {1098-0121, 1550-235X},
	url = {https://link.aps.org/doi/10.1103/PhysRevB.87.184115},
	doi = {10.1103/PhysRevB.87.184115},
	language = {en},
	number = {18},
	urldate = {2026-03-05},
	journal = {Physical Review B},
	author = {Bartók, Albert P. and Kondor, Risi and Csányi, Gábor},
	month = may,
	year = {2013},
	pages = {184115},
}

@book{siegbahn_esca_1970,
	address = {Amsterdam},
	title = {{ESCA} applied to free molecules. {By} {K}. {Siegbahn} [and others]},
	isbn = {0-7204-0160-7},
	language = {eng},
	publisher = {North-Holland Pub. Co.},
	author = {Siegbahn, Kai},
	year = {1970},
	lccn = {75475622},
	note = {Publication Title: ESCA applied to free molecules.},
}

@article{bagus_interpretation_2013,
	title = {The interpretation of {XPS} spectra: {Insights} into materials properties},
	volume = {68},
	issn = {01675729},
	shorttitle = {The interpretation of {XPS} spectra},
	url = {https://linkinghub.elsevier.com/retrieve/pii/S0167572913000125},
	doi = {10.1016/j.surfrep.2013.03.001},
	language = {en},
	number = {2},
	urldate = {2026-03-16},
	journal = {Surface Science Reports},
	author = {Bagus, Paul S. and Ilton, Eugene S. and Nelin, Connie J.},
	month = jun,
	year = {2013},
	pages = {273--304},
}

@article{seidel_valence_2016,
	title = {Valence {Electronic} {Structure} of {Aqueous} {Solutions}: {Insights} from {Photoelectron} {Spectroscopy}},
	volume = {67},
	issn = {0066-426X, 1545-1593},
	shorttitle = {Valence {Electronic} {Structure} of {Aqueous} {Solutions}},
	url = {https://www.annualreviews.org/doi/10.1146/annurev-physchem-040513-103715},
	doi = {10.1146/annurev-physchem-040513-103715},
	language = {en},
	number = {1},
	urldate = {2026-03-16},
	journal = {Annual Review of Physical Chemistry},
	author = {Seidel, Robert and Winter, Bernd and Bradforth, Stephen E.},
	month = may,
	year = {2016},
	pages = {283--305},
}

@article{ratner_surface_1983,
	title = {Surface characterization of biomaterials by electron spectroscopy for chemical analysis},
	volume = {11},
	copyright = {http://www.springer.com/tdm},
	issn = {0090-6964, 1573-9686},
	url = {http://link.springer.com/10.1007/BF02363290},
	doi = {10.1007/BF02363290},
	language = {en},
	number = {3-4},
	urldate = {2026-03-16},
	journal = {Annals of Biomedical Engineering},
	author = {Ratner, Buddy D.},
	month = may,
	year = {1983},
	pages = {313--336},
}

@article{hurisso_amino_2011,
	title = {Amino acid-based ionic liquids: using {XPS} to probe the electronic environment via binding energies},
	volume = {13},
	issn = {1463-9076, 1463-9084},
	shorttitle = {Amino acid-based ionic liquids},
	url = {https://xlink.rsc.org/?DOI=c1cp21763a},
	doi = {10.1039/c1cp21763a},
	language = {en},
	number = {39},
	urldate = {2026-03-16},
	journal = {Physical Chemistry Chemical Physics},
	author = {Hurisso, Bitu Birru and Lovelock, Kevin R. J. and Licence, Peter},
	year = {2011},
	pages = {17737},
}

@article{bagus_mechanisms_1999,
	title = {Mechanisms responsible for chemical shifts of core-level binding energies and their relationship to chemical bonding},
	volume = {100},
	copyright = {https://www.elsevier.com/tdm/userlicense/1.0/},
	issn = {03682048},
	url = {https://linkinghub.elsevier.com/retrieve/pii/S0368204899000481},
	doi = {10.1016/S0368-2048(99)00048-1},
	language = {en},
	number = {1-3},
	urldate = {2026-03-16},
	journal = {Journal of Electron Spectroscopy and Related Phenomena},
	author = {Bagus, Paul S and Illas, Francesc and Pacchioni, Gianfranco and Parmigiani, Fulvio},
	month = oct,
	year = {1999},
	pages = {215--236},
}

@article{ramakrishnan_quantum_2014,
	title = {Quantum chemistry structures and properties of 134 kilo molecules},
	volume = {1},
	issn = {2052-4463},
	url = {https://www.nature.com/articles/sdata201422},
	doi = {10.1038/sdata.2014.22},
	language = {en},
	number = {1},
	urldate = {2026-03-18},
	journal = {Scientific Data},
	author = {Ramakrishnan, Raghunathan and Dral, Pavlo O. and Rupp, Matthias and Von Lilienfeld, O. Anatole},
	month = aug,
	year = {2014},
	pages = {140022},
}

@article{butina_unsupervised_1999,
	title = {Unsupervised {Data} {Base} {Clustering} {Based} on {Daylight}'s {Fingerprint} and {Tanimoto} {Similarity}: {A} {Fast} and {Automated} {Way} {To} {Cluster} {Small} and {Large} {Data} {Sets}},
	volume = {39},
	issn = {0095-2338},
	shorttitle = {Unsupervised {Data} {Base} {Clustering} {Based} on {Daylight}'s {Fingerprint} and {Tanimoto} {Similarity}},
	url = {https://pubs.acs.org/doi/10.1021/ci9803381},
	doi = {10.1021/ci9803381},
	language = {en},
	number = {4},
	urldate = {2026-03-18},
	journal = {Journal of Chemical Information and Computer Sciences},
	author = {Butina, Darko},
	month = jul,
	year = {1999},
	pages = {747--750},
}

@misc{passaro_reducing_2023,
	title = {Reducing {SO}(3) {Convolutions} to {SO}(2) for {Efficient} {Equivariant} {GNNs}},
	url = {http://arxiv.org/abs/2302.03655},
	doi = {10.48550/arXiv.2302.03655},
	language = {en},
	urldate = {2026-03-18},
	publisher = {arXiv},
	author = {Passaro, Saro and Zitnick, C. Lawrence},
	month = jun,
	year = {2023},
	note = {arXiv:2302.03655 [cs]},
}

@misc{batatia_mace_2023,
	title = {{MACE}: {Higher} {Order} {Equivariant} {Message} {Passing} {Neural} {Networks} for {Fast} and {Accurate} {Force} {Fields}},
	shorttitle = {{MACE}},
	url = {http://arxiv.org/abs/2206.07697},
	doi = {10.48550/arXiv.2206.07697},
	language = {en},
	urldate = {2026-03-18},
	publisher = {arXiv},
	author = {Batatia, Ilyes and Kovács, Dávid Péter and Simm, Gregor N. C. and Ortner, Christoph and Csányi, Gábor},
	month = jan,
	year = {2023},
	note = {arXiv:2206.07697 [stat]},
}

@misc{geiger_e3nn_2022-1,
	title = {e3nn: {Euclidean} {Neural} {Networks}},
	shorttitle = {e3nn},
	url = {http://arxiv.org/abs/2207.09453},
	doi = {10.48550/arXiv.2207.09453},
	language = {en},
	urldate = {2026-03-18},
	publisher = {arXiv},
	author = {Geiger, Mario and Smidt, Tess},
	month = jul,
	year = {2022},
	note = {arXiv:2207.09453 [cs]},
}

@article{dick_probing_2020,
	title = {Probing the electronic structure of ether functionalised ionic liquids using {X}-ray photoelectron spectroscopy},
	volume = {22},
	url = {http://dx.doi.org/10.1039/C9CP01297D},
	doi = {10.1039/C9CP01297D},
	number = {3},
	journal = {Phys. Chem. Chem. Phys.},
	publisher = {The Royal Society of Chemistry},
	author = {Dick, Ejike J. and Fouda, Adam E. A. and Besley, Nicholas A. and Licence, Peter},
	year = {2020},
	pages = {1624--1631},
}

@article{tenorio_multi-reference_2021,
	title = {Multi-reference approach to the computation of double core-hole spectra},
	volume = {155},
	issn = {0021-9606},
	url = {https://doi.org/10.1063/5.0062130},
	doi = {10.1063/5.0062130},
	number = {13},
	journal = {The Journal of Chemical Physics},
	author = {Tenorio, Bruno Nunes Cabral and Decleva, Piero and Coriani, Sonia},
	month = oct,
	year = {2021},
	note = {\_eprint: https://pubs.aip.org/aip/jcp/article-pdf/doi/10.1063/5.0062130/16043240/131101\_1\_online.pdf},
	pages = {131101},
}

@article{prange_toward_2025,
	title = {Toward a {Machine} {Learning} {Approach} to {Interpreting} {X}-ray {Spectra} of {Trace} {Impurities} by {Converting} {XANES} to {EXAFS}},
	volume = {129},
	copyright = {https://doi.org/10.15223/policy-029},
	issn = {1089-5639, 1520-5215},
	url = {https://pubs.acs.org/doi/10.1021/acs.jpca.4c05612},
	doi = {10.1021/acs.jpca.4c05612},
	language = {en},
	number = {1},
	urldate = {2026-04-16},
	journal = {The Journal of Physical Chemistry A},
	author = {Prange, Micah P. and Govind, Niranjan and Stinis, Panos and Ilton, Eugene S. and Howard, Amanda A.},
	month = jan,
	year = {2025},
	pages = {346--355},
}

@article{mejia-rodriguez_scalable_2021,
	title = {Scalable {Molecular} {GW} {Calculations}: {Valence} and {Core} {Spectra}},
	volume = {17},
	copyright = {https://doi.org/10.15223/policy-029},
	issn = {1549-9618, 1549-9626},
	shorttitle = {Scalable {Molecular} {GW} {Calculations}},
	url = {https://pubs.acs.org/doi/10.1021/acs.jctc.1c00738},
	doi = {10.1021/acs.jctc.1c00738},
	language = {en},
	number = {12},
	urldate = {2026-04-16},
	journal = {Journal of Chemical Theory and Computation},
	author = {Mejia-Rodriguez, Daniel and Kunitsa, Alexander and Aprà, Edoardo and Govind, Niranjan},
	month = dec,
	year = {2021},
	pages = {7504--7517},
}

@article{mejia-rodriguez_basis_2022,
	title = {Basis {Set} {Selection} for {Molecular} {Core}-{Level} \textit{{GW}} {Calculations}},
	volume = {18},
	copyright = {https://doi.org/10.15223/policy-029},
	issn = {1549-9618, 1549-9626},
	url = {https://pubs.acs.org/doi/10.1021/acs.jctc.2c00247},
	doi = {10.1021/acs.jctc.2c00247},
	language = {en},
	number = {8},
	urldate = {2026-04-16},
	journal = {Journal of Chemical Theory and Computation},
	author = {Mejia-Rodriguez, Daniel and Kunitsa, Alexander and Aprà, Edoardo and Govind, Niranjan},
	month = aug,
	year = {2022},
	pages = {4919--4926},
}

@misc{fouda_2026_19689244,
  author       = {Fouda, Adam Emad Ahmed},
  title        = {AugerNet: GNN predictions of carbon 1s core-
                   electron binding energies.
                  },
  month        = apr,
  year         = 2026,
  publisher    = {Zenodo},
  version      = {v1.0.0},
  doi          = {10.5281/zenodo.19689244},
  url          = {https://doi.org/10.5281/zenodo.19689244},
}

@misc{fouda2026experimentallyaccurategraphneural,
      title={Experimentally Accurate Graph Neural Network Predictions of Core-Electron Binding Energies}, 
      author={Adam E. A. Fouda and Joshua Zhou and Rodrigo Ferreira and Patrick Phillips and Valay Agarawal and Bhavnesh Jangid and Jacob J. Wardzala and Rui Ding and Junhong Chen and Nicole Tebaldi and Phay J. Ho and Laura Gagliardi and Linda Young},
      year={2026},
      eprint={2604.27070},
      archivePrefix={arXiv},
      primaryClass={physics.chem-ph},
      url={https://arxiv.org/abs/2604.27070}, 
}

@article{doi:10.1021/acsanm.5c00100,
author = {Bouchabou, Meryem and Rives López, Juan Manuel and Román Martínez, María del Carmen and Lillo-Rodenas, María Angeles},
title = {Evaluating H2 Production by Ultraviolet-Induced Water Splitting over (Cu or Ni)-TiO2 Nanoparticle Photocatalysts},
journal = {ACS Applied Nano Materials},
volume = {8},
number = {17},
pages = {8646-8662},
year = {2025},
doi = {10.1021/acsanm.5c00100},
URL = { 
        https://doi.org/10.1021/acsanm.5c00100
},
eprint = { 
        https://doi.org/10.1021/acsanm.5c00100
}
}

@article{https://doi.org/10.1002/sia.6239,
author = {Biesinger, Mark C.},
title = {Advanced analysis of copper X-ray photoelectron spectra},
journal = {Surface and Interface Analysis},
volume = {49},
number = {13},
pages = {1325-1334},
doi = {https://doi.org/10.1002/sia.6239},
url = {https://analyticalsciencejournals.onlinelibrary.wiley.com/doi/abs/10.1002/sia.6239},
eprint = {https://analyticalsciencejournals.onlinelibrary.wiley.com/doi/pdf/10.1002/sia.6239},
year = {2017}
}

@article{SOLDEMO2024122565,
title = {Using Auger transitions as a route to determine the oxidation state of copper in high-pressure electron spectroscopy},
journal = {Surface Science},
volume = {749},
pages = {122565},
year = {2024},
issn = {0039-6028},
doi = {https://doi.org/10.1016/j.susc.2024.122565},
url = {https://www.sciencedirect.com/science/article/pii/S003960282400116X},
author = {Markus Soldemo and Fernando Garcia-Martinez and Christopher M Goodwin and Patrick Lömker and Mikhail Shipilin and Anders Nilsson and Peter Amann and Sarp Kaya and Jonas Weissenrieder}
}

@article{HENDERSON2025147578,
title = {Applications of auger electron spectroscopy in the chemical state analysis of copper and its oxides},
journal = {Journal of Electron Spectroscopy and Related Phenomena},
volume = {283},
pages = {147578},
year = {2025},
issn = {0368-2048},
doi = {https://doi.org/10.1016/j.elspec.2025.147578},
url = {https://www.sciencedirect.com/science/article/pii/S0368204825000659},
author = {Jeffrey D. Henderson and Mohammad Sabeti and Xuejie Li and Na Wang and Mehran Behazin and Mark C. Biesinger and James J. Noël and Sridhar Ramamurthy}
}

@article{FOX1977390,
title = {Solid state effects in the Auger spectrum of zinc and oxidised zinc},
journal = {Surface Science},
volume = {63},
pages = {390-402},
year = {1977},
issn = {0039-6028},
doi = {https://doi.org/10.1016/0039-6028(77)90354-5},
url = {https://www.sciencedirect.com/science/article/pii/0039602877903545},
author = {J.H Fox and J.D Nuttall and T.E Gallon}
}

@article{10.1039/c2cp22419d,
    author = {Biesinger, Mark C. and Lau, Leo W. M. and Gerson, Andrea R. and Smart, Roger St. C.},
    title = {The role of the Auger parameter in XPS studies of nickel metal, halides and oxides},
    journal = {Physical Chemistry Chemical Physics},
    volume = {14},
    number = {7},
    pages = {2434-2442},
    year = {2012},
    month = {02},
    issn = {1463-9076},
    doi = {10.1039/c2cp22419d},
    url = {https://doi.org/10.1039/c2cp22419d},
    eprint = {https://pubs.rsc.org/cp/article-pdf/14/7/2434/2577395/c2cp22419d.pdf},
}

@article{https://doi.org/10.1002/admi.202201828,
author = {Wieczorek, Alexander and Lai, Huagui and Pious, Johnpaul and Fu, Fan and Siol, Sebastian},
title = {Resolving Oxidation States and X–site Composition of Sn Perovskites through Auger Parameter Analysis in XPS},
journal = {Advanced Materials Interfaces},
volume = {10},
number = {7},
pages = {2201828},
doi = {https://doi.org/10.1002/admi.202201828},
url = {https://advanced.onlinelibrary.wiley.com/doi/abs/10.1002/admi.202201828},
eprint = {https://advanced.onlinelibrary.wiley.com/doi/pdf/10.1002/admi.202201828},
year = {2023}
}

@article{doi:10.1021/jp0564400,
author = {Hu, Chun and Lan, Yongqing and Qu, Jiuhui and Hu, Xuexiang and Wang, Aimin},
title = {Ag/AgBr/TiO2 Visible Light Photocatalyst for Destruction of Azodyes and Bacteria},
journal = {The Journal of Physical Chemistry B},
volume = {110},
number = {9},
pages = {4066-4072},
year = {2006},
doi = {10.1021/jp0564400},
    note ={PMID: 16509698},
URL = {   
        https://doi.org/10.1021/jp0564400
},
eprint = {  
        https://doi.org/10.1021/jp0564400
}
}

@article{doi:10.1021/acsomega.1c05002,
author = {Lin, Wei-Chun and Lo, Wei-Chun and Li, Jun-Xian and Huang, Pei-Chen and Wang, Man-Ying},
title = {Auger Electron Spectroscopy Analysis of the Thermally Induced Degradation of MAPbI3 Perovskite Films},
journal = {ACS Omega},
volume = {6},
number = {50},
pages = {34606-34614},
year = {2021},
doi = {10.1021/acsomega.1c05002},
URL = { 
        https://doi.org/10.1021/acsomega.1c05002
},
eprint = { 
        https://doi.org/10.1021/acsomega.1c05002
}
}

@article{WAGNER1988283,
title = {The auger parameter, its utility and advantages: a review},
journal = {Journal of Electron Spectroscopy and Related Phenomena},
volume = {47},
pages = {283-313},
year = {1988},
issn = {0368-2048},
doi = {https://doi.org/10.1016/0368-2048(88)85018-7},
url = {https://www.sciencedirect.com/science/article/pii/0368204888850187},
author = {C.D. Wagner and A. Joshi}
}

@article{MORETTI20133,
title = {The Wagner plot and the Auger parameter as tools to separate initial- and final-state contributions in X-ray photoemission spectroscopy},
journal = {Surface Science},
volume = {618},
pages = {3-11},
year = {2013},
issn = {0039-6028},
doi = {https://doi.org/10.1016/j.susc.2013.09.009},
url = {https://www.sciencedirect.com/science/article/pii/S003960281300263X},
author = {Giuliano Moretti}
}

@Article{Pelimanni2024,
author={Pelimanni, Eetu
and Fouda, Adam E. A.
and Ho, Phay J.
and Baumann, Thomas M.
and Bokarev, Sergey I.
and Fanis, Alberto De
and Dold, Simon
and Grell, Gilbert
and Ismail, Iyas
and Koulentianos, Dimitrios
and Mazza, Tommaso
and Meyer, Michael
and Piancastelli, Maria-Novella
and P{\"u}ttner, Ralph
and Rivas, Daniel E.
and Senfftleben, Bj{\"o}rn
and Simon, Marc
and Young, Linda
and Doumy, Gilles},
title={Observation of molecular resonant double-core excitation driven by intense X-ray pulses},
journal={Communications Physics},
year={2024},
month={Oct},
day={17},
volume={7},
number={1},
pages={341},
issn={2399-3650},
doi={10.1038/s42005-024-01804-5},
url={https://doi.org/10.1038/s42005-024-01804-5}
}

@article{Thompson_2024,
doi = {10.1088/1361-6455/ad7e89},
url = {https://dx.doi.org/10.1088/1361-6455/ad7e89},
year = {2024},
month = {oct},
publisher = {IOP Publishing},
volume = {57},
number = {21},
pages = {215602},
author = {Henry J Thompson and Oksana Plekan and Matteo Bonanomi and Nitish Pal and Felix Allum and Alexander D Brynes and Marcello Coreno and Sonia Coriani and Miltcho B Danailov and Piero Decleva and Alexander Demidovich and Michele Devetta and Davide Faccialà and Raimund Feifel and Ruaridh Forbes and Cesare Grazioli and David M P Holland and Paolo Piseri and Kevin C Prince and Daniel Rolles and Michael S Schuurman and Alberto Simoncig and Richard J Squibb and Bruno N C Tenorio and Caterina Vozzi and Marco Zangrando and Carlo Callegari and Russell S Minns and Michele Di Fraia},
title = {Time-resolved Auger–Meitner spectroscopy of the photodissociation dynamics of CS2},
journal = {Journal of Physics B: Atomic, Molecular and Optical Physics}
}

@Article{D1CP00623A,
author ="Kissin, Yoel and Ruberti, Marco and Kolorenč, Přemysl and Averbukh, Vitali",
title  ="Attosecond pump–attosecond probe spectroscopy of Auger decay",
journal  ="Physical Chemistry Chemical Physics",
year  ="2021",
volume  ="23",
issue  ="21",
pages  ="12376-12386",
publisher  ="The Royal Society of Chemistry",
doi  ="10.1039/D1CP00623A",
url  ="http://dx.doi.org/10.1039/D1CP00623A"
}

@article{
doi:10.1126/science.abj2096,
author = {Siqi Li  and Taran Driver  and Philipp Rosenberger  and Elio G. Champenois  and Joseph Duris  and Andre Al-Haddad  and Vitali Averbukh  and Jonathan C. T. Barnard  and Nora Berrah  and Christoph Bostedt  and Philip H. Bucksbaum  and Ryan N. Coffee  and Louis F. DiMauro  and Li Fang  and Douglas Garratt  and Averell Gatton  and Zhaoheng Guo  and Gregor Hartmann  and Daniel Haxton  and Wolfram Helml  and Zhirong Huang  and Aaron C. LaForge  and Andrei Kamalov  and Jonas Knurr  and Ming-Fu Lin  and Alberto A. Lutman  and James P. MacArthur  and Jon P. Marangos  and Megan Nantel  and Adi Natan  and Razib Obaid  and Jordan T. O’Neal  and Niranjan H. Shivaram  and Aviad Schori  and Peter Walter  and Anna Li Wang  and Thomas J. A. Wolf  and Zhen Zhang  and Matthias F. Kling  and Agostino Marinelli  and James P. Cryan },
title = {Attosecond coherent electron motion in Auger-Meitner decay},
journal = {Science},
volume = {375},
number = {6578},
pages = {285-290},
year = {2022},
doi = {10.1126/science.abj2096},
URL = {https://www.science.org/doi/abs/10.1126/science.abj2096},
eprint = {https://www.science.org/doi/pdf/10.1126/science.abj2096}
}

@Article{Driver2024,
author={Driver, Taran
and Mountney, Miles
and Wang, Jun
and Ortmann, Lisa
and Al-Haddad, Andre
and Berrah, Nora
and Bostedt, Christoph
and Champenois, Elio G.
and DiMauro, Louis F.
and Duris, Joseph
and Garratt, Douglas
and Glownia, James M.
and Guo, Zhaoheng
and Haxton, Daniel
and Isele, Erik
and Ivanov, Igor
and Ji, Jiabao
and Kamalov, Andrei
and Li, Siqi
and Lin, Ming-Fu
and Marangos, Jon P.
and Obaid, Razib
and O'Neal, Jordan T.
and Rosenberger, Philipp
and Shivaram, Niranjan H.
and Wang, Anna L.
and Walter, Peter
and Wolf, Thomas J. A.
and W{\"o}rner, Hans Jakob
and Zhang, Zhen
and Bucksbaum, Philip H.
and Kling, Matthias F.
and Landsman, Alexandra S.
and Lucchese, Robert R.
and Emmanouilidou, Agapi
and Marinelli, Agostino
and Cryan, James P.},
title={Attosecond delays in X-ray molecular ionization},
journal={Nature},
year={2024},
month={Aug},
day={01},
volume={632},
number={8026},
pages={762-767},
issn={1476-4687},
doi={10.1038/s41586-024-07771-9},
url={https://doi.org/10.1038/s41586-024-07771-9}
}

@article{doi:10.1080/09553002.2020.1831706,
author = {Roger W. Howell},
title = {Advancements in the use of Auger electrons in science and medicine during the period 2015–2019},
journal = {International Journal of Radiation Biology},
volume = {99},
number = {1},
pages = {2--27},
year = {2023},
publisher = {Taylor \& Francis},
doi = {10.1080/09553002.2020.1831706},
    note ={PMID: 33021416},
URL = {  
        https://doi.org/10.1080/09553002.2020.1831706
},
eprint = {    
        https://doi.org/10.1080/09553002.2020.1831706
}
}

@Article{Ku2019,
author={Ku, Anthony
and Facca, Valerie J.
and Cai, Zhongli
and Reilly, Raymond M.},
title={Auger electrons for cancer therapy -- a review},
journal={EJNMMI Radiopharmacy and Chemistry},
year={2019},
month={Oct},
day={11},
volume={4},
number={1},
pages={27},
issn={2365-421X},
doi={10.1186/s41181-019-0075-2},
url={https://doi.org/10.1186/s41181-019-0075-2}
}

@article{Stollenwerk_2025,
doi = {10.1088/1367-2630/add6cd},
url = {https://doi.org/10.1088/1367-2630/add6cd},
year = {2025},
month = {may},
publisher = {IOP Publishing},
volume = {27},
number = {5},
pages = {053901},
author = {Stollenwerk, Patrick R and Southworth, Stephen H and Granato, Francesco and Renne, Amy and Mustapha, Brahim and Bailey, Kevin G and Mueller, Peter and Nolen, Jerry and O’Connor, Thomas P and Xie, Junqi and Young, Linda and Dietrich, Matthew R},
title = {The Auger Radioisotope Microscope: an instrument for characterization of Auger electron multiplicities and energy distributions},
journal = {New Journal of Physics}
}

@article{doi:10.1021/acs.jpca.5c01789,
author = {Fouda, Adam E. A. and Jangid, Bhavnesh and Pelimanni, Eetu and Southworth, Stephen H. and Ho, Phay J. and Gagliardi, Laura and Young, Linda},
title = {Computation of Auger Electron Spectra in Organic Molecules with Multiconfiguration Pair-Density Functional Theory},
journal = {The Journal of Physical Chemistry A},
volume = {129},
number = {36},
pages = {8419-8431},
year = {2025},
doi = {10.1021/acs.jpca.5c01789},
    note ={PMID: 40879544},
URL = {    
        https://doi.org/10.1021/acs.jpca.5c01789
},
eprint = { 
        https://doi.org/10.1021/acs.jpca.5c01789
}
}

@article{doi:10.1063/1.1386414,
	author = {Deleuze,Michael S. and Trofimov,Alexander B. and Cederbaum, Lorenz S.},
	doi = {10.1063/1.1386414},
	eprint = {https://doi.org/10.1063/1.1386414},
	journal = {The Journal of Chemical Physics},
	number = {13},
	pages = {5859-5882},
	title = {Valence one-electron and shake-up ionization bands of polycyclic aromatic hydrocarbons. I. Benzene, naphthalene, anthracene, naphthacene, and pentacene},
	url = {https://doi.org/10.1063/1.1386414},
	volume = {115},
	year = {2001}}

@article{MITANI2003103,
title = {Theoretical molecular Auger spectra with electron population analysis},
journal = {Journal of Electron Spectroscopy and Related Phenomena},
volume = {128},
number = {2},
pages = {103-117},
year = {2003},
issn = {0368-2048},
doi = {https://doi.org/10.1016/S0368-2048(02)00270-0},
url = {https://www.sciencedirect.com/science/article/pii/S0368204802002700},
author = {Masaki Mitani and Osamu Takahashi and Ko Saito and Suehiro Iwata}
}

@article{10.1063/1.2166234,
    author = {Takahashi, Osamu and Odelius, Michael and Nordlund, Dennis and Nilsson, Anders and Bluhm, Hendrik and Pettersson, Lars G. M.},
    title = "{Auger decay calculations with core-hole excited-state molecular-dynamics simulations of water}",
    journal = {The Journal of Chemical Physics},
    volume = {124},
    number = {6},
    pages = {064307},
    year = {2006},
    month = {02},
    issn = {0021-9606},
    doi = {10.1063/1.2166234},
    url = {https://doi.org/10.1063/1.2166234},
    eprint = {https://pubs.aip.org/aip/jcp/article-pdf/doi/10.1063/1.2166234/14038028/064307\_1\_online.pdf},
}

@article{doi:10.1021/acs.jpclett.3c03611,
author = {Fouda, A. E. A. and Lindblom, V. and Southworth, S. H. and Doumy, G. and Ho, P. J. and Young, L. and Cheng, L. and Sorensen, S. L.},
title = {Influence of Selective Carbon 1s Excitation on Auger–Meitner Decay in the ESCA Molecule},
journal = {The Journal of Physical Chemistry Letters},
volume = {15},
number = {16},
pages = {4286-4293},
year = {2024},
doi = {10.1021/acs.jpclett.3c03611},
    note ={PMID: 38608168},
URL = { 
    
        https://doi.org/10.1021/acs.jpclett.3c03611
},
eprint = { 
    
        https://doi.org/10.1021/acs.jpclett.3c03611
}
}

@Article{D3CP01746J,
author ="de Moura, C. E. V. and Laurent, J. and Bozek, J. and Briant, M. and Çarçabal, P. and Cubaynes, D. and Shafizadeh, N. and Simon, M. and Soep, B. and Püttner, R. and et al.",
title  ="Experimental and theoretical study of resonant core-hole spectroscopies of gas-phase free-base phthalocyanine",
journal  ="Physical Chemistry Chemical Physics",
year  ="2023",
volume  ="25",
issue  ="22",
pages  ="15555-15566",
publisher  ="The Royal Society of Chemistry",
doi  ="10.1039/D3CP01746J",
url  ="http://dx.doi.org/10.1039/D3CP01746J"}

@Article{C7CP02345F,
author ="Banks, H. I. B. and Little, D. A. and Tennyson, J. and Emmanouilidou, A.",
title  ="Interaction of molecular nitrogen with free-electron-laser radiation",
journal  ="Physical Chemistry Chemical Physics",
year  ="2017",
volume  ="19",
issue  ="30",
pages  ="19794-19806",
publisher  ="The Royal Society of Chemistry",
doi  ="10.1039/C7CP02345F",
url  ="http://dx.doi.org/10.1039/C7CP02345F"}

@article{10.1063/1.3700233,
    author = {Inhester, L. and Burmeister, C. F. and Groenhof, G. and Grubmüller, H.},
    title = "{Auger spectrum of a water molecule after single and double core ionization}",
    journal = {The Journal of Chemical Physics},
    volume = {136},
    number = {14},
    pages = {144304},
    year = {2012},
    month = {04},
    issn = {0021-9606},
    doi = {10.1063/1.3700233},
    url = {https://doi.org/10.1063/1.3700233},
    eprint = {https://pubs.aip.org/aip/jcp/article-pdf/doi/10.1063/1.3700233/15451343/144304\_1\_online.pdf},
}

@article{10.1063/1.3526026,
    author = {Demekhin, Ph. V. and Ehresmann, A. and Sukhorukov, V. L.},
    title = "{Single center method: A computational tool for ionization and electronic excitation studies of molecules}",
    journal = {The Journal of Chemical Physics},
    volume = {134},
    number = {2},
    pages = {024113},
    year = {2011},
    month = {01},
    issn = {0021-9606},
    doi = {10.1063/1.3526026},
    url = {https://doi.org/10.1063/1.3526026},
    eprint = {https://pubs.aip.org/aip/jcp/article-pdf/doi/10.1063/1.3526026/13040006/024113\_1\_online.pdf},
}

@article{PhysRevA.80.063425,
  title = {Interference effects during the Auger decay of the ${\text{C}}^{\ensuremath{\ast}}\text{O}$ $1{s}^{\ensuremath{-}1}{\ensuremath{\pi}}^{\ensuremath{\ast}}$ resonance studied by angular distribution of the ${\text{CO}}^{+}(A)$ photoelectrons and polarization analysis of the ${\text{CO}}^{+}(A\text{\ensuremath{-}}X)$ fluorescence},
  author = {Demekhin, Ph. V. and Petrov, I. D. and Sukhorukov, V. L. and Kielich, W. and Reiss, P. and Hentges, R. and Haar, I. and Schmoranzer, H. and Ehresmann, A.},
  journal = {Physical Review A},
  volume = {80},
  issue = {6},
  pages = {063425},
  numpages = {12},
  year = {2009},
  month = {Dec},
  publisher = {American Physical Society},
  doi = {10.1103/PhysRevA.80.063425},
  url = {https://link.aps.org/doi/10.1103/PhysRevA.80.063425}
}

@Article{Demekhin2007,
author={Demekhin, Ph. V.
and Omel'yanenko, D. V.
and Lagutin, B. M.
and Sukhorukov, V. L.
and Werner, L.
and Ehresmann, A.
and Schartner, K.-H.
and Schmoranzer, H.},
title={Investigation of photoionization and photodissociation of an oxygen molecule by the method of coupled differential equations},
journal={Optics and Spectroscopy},
year={2007},
month={Mar},
day={01},
volume={102},
number={3},
pages={318-329},
issn={1562-6911},
doi={10.1134/S0030400X07030022},
url={https://doi.org/10.1134/S0030400X07030022}
}

@article{PhysRevA.45.318,
  title = {Molecular scattering wave functions for Auger decay rates: The Auger spectrum of hydrogen fluoride},
  author = {Z\"ahringer, K. and Meyer, H.-D. and Cederbaum, L. S.},
  journal = {Physical Review A},
  volume = {45},
  issue = {1},
  pages = {318--328},
  numpages = {0},
  year = {1992},
  month = {Jan},
  publisher = {American Physical Society},
  doi = {10.1103/PhysRevA.45.318},
  url = {https://link.aps.org/doi/10.1103/PhysRevA.45.318}
}

@article{HIGASHI1982377,
title = {Calculations of the Auger transition rates in molecules. I. Effect of the nonspherical potential: Application to CH4},
journal = {Chemical Physics},
volume = {68},
number = {3},
pages = {377-382},
year = {1982},
issn = {0301-0104},
doi = {https://doi.org/10.1016/0301-0104(82)87045-6},
url = {https://www.sciencedirect.com/science/article/pii/0301010482870456},
author = {M. Higashi and E. Hiroike and T. Nakajima}
}

@article{PhysRevA.19.1649,
  title = {Calculated Auger transition rates for HF},
  author = {Faegri, K. and Kelly, H. P.},
  journal = {Physical Review A},
  volume = {19},
  issue = {4},
  pages = {1649--1655},
  numpages = {0},
  year = {1979},
  month = {Apr},
  publisher = {American Physical Society},
  doi = {10.1103/PhysRevA.19.1649},
  url = {https://link.aps.org/doi/10.1103/PhysRevA.19.1649}
}

@article{10.1063/1.4919794,
    author = {Hao, Yajiang and Inhester, Ludger and Hanasaki, Kota and Son, Sang-Kil and Santra, Robin},
    title = "{Efficient electronic structure calculation for molecular ionization dynamics at high x-ray intensity}",
    journal = {Structural Dynamics},
    volume = {2},
    number = {4},
    pages = {041707},
    year = {2015},
    month = {05},
    issn = {2329-7778},
    doi = {10.1063/1.4919794},
    url = {https://doi.org/10.1063/1.4919794},
    eprint = {https://pubs.aip.org/aca/sdy/article-pdf/doi/10.1063/1.4919794/13755898/041707\_1\_online.pdf},
}

@article{PhysRevA.94.023422,
  title = {X-ray multiphoton ionization dynamics of a water molecule irradiated by an x-ray free-electron laser pulse},
  author = {Inhester, Ludger and Hanasaki, Kota and Hao, Yajiang and Son, Sang-Kil and Santra, Robin},
  journal = {Physical Review A},
  volume = {94},
  issue = {2},
  pages = {023422},
  numpages = {8},
  year = {2016},
  month = {Aug},
  publisher = {American Physical Society},
  doi = {10.1103/PhysRevA.94.023422},
  url = {https://link.aps.org/doi/10.1103/PhysRevA.94.023422}
}

@article{TRAVNIKOVA200967,
title = {Assignment of the L2,3VV normal Auger decay spectrum of Cl2 by ab initio calculations},
journal = {Chemical Physics Letters},
volume = {474},
number = {1},
pages = {67-73},
year = {2009},
issn = {0009-2614},
doi = {https://doi.org/10.1016/j.cplett.2009.04.058},
url = {https://www.sciencedirect.com/science/article/pii/S0009261409004795},
author = {Oksana Travnikova and Reinhold F. Fink and Antti Kivimäki and Denis Céolin and Zhuo Bao and Maria Novella Piancastelli}
}

@article{FINK1995295,
title = {Theoretical autoionization spectra of 1s \(\to\) \(\pi^\ast\) excited N2 and N2O},
journal = {Journal of Electron Spectroscopy and Related Phenomena},
volume = {76},
pages = {295-300},
year = {1995},
note = {Proceedings of the Sixth International Conference on Electron Spectroscopy},
issn = {0368-2048},
doi = {https://doi.org/10.1016/0368-2048(95)02469-7},
url = {https://www.sciencedirect.com/science/article/pii/0368204895024697},
author = {Reinhold Fink}
}

@Article{Larkins1990,
author={Larkins, F. P.
and Tulea, L. C.
and Chelkowska, E. Z.},
title={Auger Electron Spectra of Molecules: The First Row Hydrides},
journal={Australian Journal of Physics},
year={1990},
volume={43},
number={5},
pages={625-640},
doi={10.1071/PH900625},
url={https://doi.org/10.1071/PH900625}
}

@article{JENNISON1980435,
title = {The calculation of molecular and cluster auger spectra},
journal = {Chemical Physics Letters},
volume = {69},
number = {3},
pages = {435-440},
year = {1980},
issn = {0009-2614},
doi = {https://doi.org/10.1016/0009-2614(80)85099-8},
url = {https://www.sciencedirect.com/science/article/pii/0009261480850998},
author = {Dwight R. Jennison}
}

@article{SIEGBAHN1975330,
title = {The Auger electron spectrum of water vapour},
journal = {Chemical Physics Letters},
volume = {35},
number = {3},
pages = {330-335},
year = {1975},
issn = {0009-2614},
doi = {https://doi.org/10.1016/0009-2614(75)85615-6},
url = {https://www.sciencedirect.com/science/article/pii/0009261475856156},
author = {H. Siegbahn and L. Asplund and P. Kelfve}
}

@article{10.1063/5.0036976,
    author = {Skomorowski, Wojciech and Krylov, Anna I.},
    title = "{Feshbach–Fano approach for calculation of Auger decay rates using equation-of-motion coupled-cluster wave functions. I. Theory and implementation}",
    journal = {The Journal of Chemical Physics},
    volume = {154},
    number = {8},
    pages = {084124},
    year = {2021},
    month = {02},
    issn = {0021-9606},
    doi = {10.1063/5.0036976},
    url = {https://doi.org/10.1063/5.0036976},
    eprint = {https://pubs.aip.org/aip/jcp/article-pdf/doi/10.1063/5.0036976/15587589/084124\_1\_online.pdf},
}

@article{10.1063/1.2126976,
    author = {Averbukh, Vitali and Cederbaum, Lorenz S.},
    title = "{Ab initio calculation of interatomic decay rates by a combination of the Fano ansatz, Green’s-function methods, and the Stieltjes imaging technique}",
    journal = {The Journal of Chemical Physics},
    volume = {123},
    number = {20},
    pages = {204107},
    year = {2005},
    month = {11},
    issn = {0021-9606},
    doi = {10.1063/1.2126976},
    url = {https://doi.org/10.1063/1.2126976},
    eprint = {https://pubs.aip.org/aip/jcp/article-pdf/doi/10.1063/1.2126976/15376746/204107\_1\_online.pdf},
}

@article{LIEGENER1982188,
title = {Auger spectra by the Green's function method},
journal = {Chemical Physics Letters},
volume = {90},
number = {3},
pages = {188-192},
year = {1982},
issn = {0009-2614},
doi = {https://doi.org/10.1016/0009-2614(82)80022-5},
url = {https://www.sciencedirect.com/science/article/pii/0009261482800225},
author = {Christoph-Maria Liegener}
}

@article{SCHIMMELPFENNIG1995173,
title = {Ab initio calculation of transition rates for autoionization: the Auger spectra of HF and F$^-$},
journal = {Journal of Electron Spectroscopy and Related Phenomena},
volume = {74},
number = {3},
pages = {173-186},
year = {1995},
issn = {0368-2048},
doi = {https://doi.org/10.1016/0368-2048(95)02373-9},
url = {https://www.sciencedirect.com/science/article/pii/0368204895023739},
author = {B. Schimmelpfennig and B.M. Nestmann and S.D. Peyerimhoff}
}

@article{BSchimmelpfennig_1992,
doi = {10.1088/0953-4075/25/6/013},
url = {https://dx.doi.org/10.1088/0953-4075/25/6/013},
year = {1992},
month = {mar},
publisher = {},
volume = {25},
number = {6},
pages = {1217},
author = {B Schimmelpfennig and  B Nestmann and  S D Peyerimhoff},
title = {Ab initio calculation of partial linewidths in the Auger decay of K-shell excited HCl},
journal = {Journal of Physics B: Atomic, Molecular and Optical Physics}
}

@article{10.1063/1.1316046,
    author = {Carravetta, V. and Ågren, H. and Vahtras, O. and Jensen, H. J. Aa.},
    title = "{Ab initio calculations of molecular resonant photoemission spectra}",
    journal = {The Journal of Chemical Physics},
    volume = {113},
    number = {18},
    pages = {7790-7798},
    year = {2000},
    month = {11},
    issn = {0021-9606},
    doi = {10.1063/1.1316046},
    url = {https://doi.org/10.1063/1.1316046},
    eprint = {https://pubs.aip.org/aip/jcp/article-pdf/113/18/7790/19287894/7790\_1\_online.pdf},
}

@InProceedings{Kendall_2018_CVPR,
author = {Kendall, Alex and Gal, Yarin and Cipolla, Roberto},
title = {Multi-Task Learning Using Uncertainty to Weigh Losses for Scene Geometry and Semantics},
booktitle = {Proceedings of the IEEE Conference on Computer Vision and Pattern Recognition (CVPR)},
month = {June},
year = {2018}
}

@article{EGELHOFF1987253,
title = {Core-level binding-energy shifts at surfaces and in solids},
journal = {Surface Science Reports},
volume = {6},
number = {6},
pages = {253-415},
year = {1987},
issn = {0167-5729},
doi = {https://doi.org/10.1016/0167-5729(87)90007-0},
url = {https://www.sciencedirect.com/science/article/pii/0167572987900070},
author = {W.F. Egelhoff}
}

@Article{Li2024,
author={Li, Wei-Hong
and Liu, Xialei
and Bilen, Hakan},
title={Universal Representations: A Unified Look at Multiple Task and Domain Learning},
journal={International Journal of Computer Vision},
year={2024},
month={May},
day={01},
volume={132},
number={5},
pages={1521-1545},
issn={1573-1405},
doi={10.1007/s11263-023-01931-6},
url={https://doi.org/10.1007/s11263-023-01931-6}
}

@InProceedings{Zhang_2023_CVPR,
    author    = {Zhang, Weixia and Zhai, Guangtao and Wei, Ying and Yang, Xiaokang and Ma, Kede},
    title     = {Blind Image Quality Assessment via Vision-Language Correspondence: A Multitask Learning Perspective},
    booktitle = {Proceedings of the IEEE/CVF Conference on Computer Vision and Pattern Recognition (CVPR)},
    month     = {June},
    year      = {2023},
    pages     = {14071-14081}
}

@inproceedings{10.1145/1390156.1390177, author = {Collobert, Ronan and Weston, Jason}, title = {A unified architecture for natural language processing: deep neural networks with multitask learning}, year = {2008}, isbn = {9781605582054}, publisher = {Association for Computing Machinery}, address = {New York, NY, USA}, url = {https://doi.org/10.1145/1390156.1390177}, doi = {10.1145/1390156.1390177}, booktitle = {Proceedings of the 25th International Conference on Machine Learning}, pages = {160–167}, numpages = {8}, location = {Helsinki, Finland}, series = {ICML '08} }

@INPROCEEDINGS{6639081,
  author={Huang, Jui-Ting and Li, Jinyu and Yu, Dong and Deng, Li and Gong, Yifan},
  booktitle={2013 IEEE International Conference on Acoustics, Speech and Signal Processing}, 
  title={Cross-language knowledge transfer using multilingual deep neural network with shared hidden layers}, 
  year={2013},
  volume={},
  number={},
  pages={7304-7308},
  doi={10.1109/ICASSP.2013.6639081}}

@article{doi:10.1021/acs.jcim.1c00646,
author = {Tan, Zheng and Li, Yan and Shi, Weimei and Yang, Shiqing},
title = {A Multitask Approach to Learn Molecular Properties},
journal = {Journal of Chemical Information and Modeling},
volume = {61},
number = {8},
pages = {3824-3834},
year = {2021},
doi = {10.1021/acs.jcim.1c00646},
    note ={PMID: 34289687},

URL = { 
    
        https://doi.org/10.1021/acs.jcim.1c00646
    
    

},
eprint = { 
    
        https://doi.org/10.1021/acs.jcim.1c00646
    
    

}

}
\end{document}